\pdfoutput=1

\documentclass[onecolumn]{article}
\usepackage{ulem}  
\usepackage{graphicx}
\usepackage{color}
\usepackage[hidelinks]{hyperref}
\usepackage{xcolor}

\newcommand{\f}{\begin{equation}}
\newcommand{\ff}{\end{equation}}
\usepackage{amsthm}
\newtheorem{definition}{Definition}
\begin{document}

\title{Biocosmology: Towards the birth of a new science}

\author{Marina Cort\^es,$^{1,2}$ Stuart A.\ Kauffman,$^{3}$ Andrew R.\ Liddle,$^{1,2}$ and Lee Smolin$^1$\\~\\
$^{1}$Perimeter Institute for Theoretical Physics, 31 Caroline Street North, \\ Waterloo, Ontario N2L 2Y5, Canada\\
$^2$Instituto de Astrof\'{\i}sica e Ci\^{e}ncias do Espa\c{c}o, Universidade de Lisboa,\\
Faculdade de Ci\^{e}ncias, Campo Grande, PT1749-016 Lisboa, Portugal\\
$^3$Institute for Systems Biology, Seattle, WA 98109, USA
}


    \maketitle
    \begin{abstract}
    \noindent
Cosmologists wish to explain how our Universe, in all its complexity, could ever have come about. For that, we assess the number of states in our Universe now. This plays the role of entropy in thermodynamics of the Universe, and reveals the magnitude of the problem of initial conditions to be solved. The usual budgeting accounts for gravity, thermal motions, and finally the vacuum energy whose entropy, given by the Bekenstein bound, dominates the entropy budget today. There is however one number which we have not accounted for: the number of states in our complex biosphere. What is the entropy of life and is it sizeable enough to need to be accounted for at the Big Bang?  Building on emerging ideas within theoretical biology, we show that the configuration space of living systems, unlike that of their fundamental physics counterparts, can grow rapidly in response to emerging biological complexity. A model for this expansion is provided through combinatorial innovation by the Theory of the Adjacent Possible (TAP) and its corresponding TAP equation, whose solutions we investigate, confirming the possibility of rapid state-pace growth. While the results of this work remain far from being firmly established, the evidence we provide is many-fold and strong. The implications are far-reaching, and open a variety of lines for future investigation, a new scientific field we term \textit{biocosmology}. In particular the relationship between the information content in life and the information content in the Universe may need to be rebuilt from scratch. 
\end{abstract}
 \vspace*{30pt}


\tableofcontents

\clearpage

\section{Introduction}

\begin{flushright}{\it In biology physicists are but observers, in puzzlement or wonder.} \\
\end{flushright}
This is a tale of initial conditions in cosmology, in the sense that it was by thinking about the problem of the initial state to the Universe that we were led to the topics addressed in this paper.

To a cosmologist, the problem of initial conditions is the question of how big is the entropy of the Universe. This in turn measures the specialness of our Universe's initial conditions, and demands a cosmological explanation. To a life scientist, the initial low entropy is an essential ingredient to enable life in all its myriad forms to exist, on Earth mostly by exploiting low-entropy photons from the Sun to fuel its processes before radiating back higher-entropy ones.

Both cosmology and the life sciences are closely tied to irreversibility. In cosmology we want to explain why the set of initial conditions has (close to) measure zero in the space of the possible configurations we find when we estimate the entropy the Universe has today. Assessing just how close to zero is our job. However, the accepted value for the Universe's entropy has been changing in recent decades, with the most recent value attributed to the domination of the entropy contributed by the vacuum energy --- we recount this evolution in Section~\ref{inventory}. This inspires us to revisit the techniques whereby we estimate the size of the problem at the Big Bang.

Specifically, our goal is to investigate whether biological systems can, via the sizes of their Hilbert spaces, make a sizeable contribution to the information content of the Universe. A strictly-reductionist viewpoint argues this is not at all possible: living beings are no more than, and can be fully understood in terms of, the fundamental particles they are composed of. The traditional argument states that we could eventually derive our living biosphere of today if we had powerful enough computers and integrated the equations of motion for all the fundamental particles of the Standard Model starting from the onset of the Universe, 13.8 billion years ago. It is commonly taken for granted that the impossibility of carrying out such an integration is only practical, and not {\it in principle}. 

However, there are many indications that biological systems cannot be understood, nor derived, in such a context. The argument expresses the view that living systems exhibit such a complex level of organization that they surpass the explanatory power of the method of pure physics \cite{rosen,ulanowicz,Kbeyond,nicholson,Ellisem}. According to such a view, even complex systems much more elementary than our biosphere can exhibit strongly-emergent phenomena. Several prominent authors have stated that even within physics such strongly-emergent phenomena should be regarded as every bit as fundamental as the elementary phenomena of particle physics and field theory \cite{anderson,leggett,laughlin}.

This article is the first of a series that explores possible two-way impacts between biology and cosmology, a topic we name {\it biocosmology}. 

Here we consider the possible impacts on cosmology of adopting modelling inspired by new trends in theoretical biology, which emphasise an unpredictable growth of the biological state space through emergent phenomena. One particular model of this is combinatorial innovation, an implementation of the Theory of the Adjacent Possible \cite{TAP} that has been proposed by one of the current authors (S.A.K.), and which can yield extraordinary growth rates of the Hilbert spaces describing living systems. At its simplest, this leads us to a reassessment of the role of biological systems in the cosmic information budget. At its most speculative, it invites a connection between the appearance of biological systems and of cosmic dark energy. Between these extremes lie many fruitful avenues of investigation impacting in both directions.

For cosmology specifically, the impact of the bridge between biology and cosmology might affect the measure of the set of boundary conditions at big bang singularity. The extreme unlikelihood of such initial values continues to be one of the most challenging and pressing questions in physics. What led to our Universe originating in such a Boltzmann-suppressed corner of configuration space? The precise quantification of the exquisite suppression of these initial conditions, and its possible re-evaluation stemming from recent developments in theoretical biology,  is the subject of this paper.

Note on terminology:  We will on occasion refer to `entropy', `states', and even `degrees of freedom' as applied to biological systems (and more generally to non-equilibrium systems). This is not because we believe that these are as yet unambiguously defined, but because these pieces of physics terminology capture the properties that we seek to elucidate in biological systems. The bridge we are building will be strengthened whenever these foundations are improved. We will often enclose those terms in quotes to remind the reader that rigorous definitions do not exist at this point.

\section{The entropy inventory}
\label{inventory}

How much information does the Universe hold, and where does it reside? We begin by carrying out an inventory of the different contributions to the Universe's entropy $S$, which is the first step towards solution of the problem of boundary conditions. The value of the entropy leads us, via Boltzmann's expression, to the number of current microstates and reveals the price to be paid in order to explain the likelihood of our world at the Big Bang. 

This price has been changing in recent years.  The entropy of the various ingredients of the physical Universe has been analyzed in detail by Egan \& Lineweaver \cite{EL}, who follow a number of seminal but more qualitative works \cite{KT,Frautschi,Penrose,Framptonetal}. We summarize and update their findings here.

\subsection{Material content}
 
We start with the particle microstates of the Universe, whose entropy (neglecting gravitational effects) comes from standard thermodynamics \cite{KT}. Although subdominant in total energy density today, the relativistic species still dominate the entropy as they are so numerous. Roughly equal contributions come from the photons of the cosmic microwave background\footnote{Eq.~(13) of Ref.~\cite{EL} has a typo; it is the entry in their Table 1 which is correct.} and from the combined three species of neutrino, each giving an entropy within the observable Universe of about $5 \times 10^{89}$ and hence $S_{\rm relativistic} \simeq  10^{90}$, where henceforth we work in natural units where the Boltzmann constant $k_{\rm B}$, gravitational constant $G$, speed of light $c$, and Planck constant $\hbar$ are all set to unity. Vopson has recently obtained essentially the same result expressed within the language of information theory \cite{vopson}. Somewhat lesser amounts come from thermal gravitons (presuming they exist) and stellar-emitted photons. The dark matter contribution is highly uncertain due to its unknown nature but is plausibly comparable to or a few orders of magnitude less than that of the photons, while the baryons that we are made of trail in a distant last. 

So, the total entropy of the particles in the observable Universe is about $10^{90}$ and the number of possible configurations is its exponential, $\exp{(10^{90})}$. But if that were the total number of microstates in the Universe today we would simply see a cold plasma, as reflected in the beautiful sky images of the {\it Planck} satellite \cite{Planck}, and not the richness of structure of the cosmic web captured in surveys like the Sloan Digital Sky Survey \cite{SDSSweb} and in large-volume simulations such as Illustris \cite{Illustris}.

\subsection{The black hole entropy}

The previous discussion included only the thermal motions of particles. To make the Universe we see today we need to switch on gravity. No one knows the entropic content in gravity except in one case, that of black holes. These appear to proliferate throughout the Universe, including supermassive black holes at the centres of galactic halos such as the magnificent $6 \times 10^9$ Solar mass black hole recently imaged by the Event Horizon Telescope within the M87 galaxy \cite{EHT}. The Bekenstein--Hawking entropy formula \cite{Bekenstein,Hawking}
\begin{equation}
S_{\rm BH} = \frac{k_{\rm B} G}{c \hbar} \, 4\pi M^2 \,,
\end{equation}
attributes black holes with a large entropy which is proportional to the square their mass $M$ (i.e.\ proportional to the horizon area, indeed up to a constant of order unity equal to the horizon area in Planck units). Frautschi \cite{Frautschi} observed that the entropy of such black holes completely dominates over the entropy of radiation, a conclusion later firmed up by Penrose \cite{Penrose}. Egan and Lineweaver \cite{EL} integrate over an observationally-motivated black hole mass spectrum, finding that the peak contribution comes from black holes of mass around $10^9$ Solar masses, to obtain $S_{\rm BH} = 3 \times 10^{104}$ within the observable Universe.\footnote{Their result updated the $S_{\rm BH} \sim 10^{101}$ estimate found by Penrose \cite{Penrose}, who assumed our own galaxy's supermassive black hole was typical. The higher-than-expected M87 black hole mass inferred from the Event Horizon Telescope observations suggests the entropy may need to be revised up further by a factor of a few, as its existence seems incompatible with the black-hole mass function used in Ref.~\cite{EL}.}

The entropy of stellar-mass black holes from the endpoint of stellar evolution lies roughly at the geometric mean of the supermassive black holes and the particles. There is also the possibility that even the supermassive black hole contribution could be beaten if the dark matter is comprised of intermediate-mass black holes of masses up to $10^5$ Solar masses \cite{frampton}, which is permitted by observations but not motivated theoretically.

If the above is correct then the problem at the Big Bang is explaining a likelihood which is one part in  $e^{10^{104}}$ (note that this is $e$ to the $10^{104}$!). In Penrose's famous picture (Ref.~\cite{Penrose}, Figure 27.21) he depicts this most exquisite fine tuning, where a deity must place a pin at this accuracy to create a Universe as special as ours.

So, are we now done counting? Not yet.

\subsection{The vacuum entropy}

Looking at the expansion history of the Universe there is a gentle curve upwards in the late Universe. That is due to the vacuum or dark energy that we can denote $\Lambda$, which dominates the energy density budget today and induces cosmic acceleration. The upturn looks innocuous, but by generating a cosmic event horizon it dominates the current entropy budget \cite{gibhawk}. As first emphasised by Davies \cite{daviesent}, the corresponding entropy is again given by the Bekenstein--Hawking area formula, now yielding\footnote{Technically, this is the entropy of the pure de Sitter space that is our asymptotic future; Egan and Lineweaver \cite{EL} show how to obtain the slightly smaller value appropriate to the present state of the Universe.}
\begin{equation}
S_{\Lambda}=\frac{k_{\rm B} c^3}{G \hbar} \, \frac{3\pi }{\Lambda} \simeq 3 \times 10^{122} \,,
\end{equation}
within the cosmic event horizon \cite{EL}, whose volume is currently about $1/25$ that of the observable Universe.  This corresponds to the holographic bound \cite{holo} and via the Boltzmann formula this corresponds to a total number of possible configurations $N_{\Lambda}=\exp\left(10^{124}\right)$ in the observable Universe. In turn this means that the going rate for explaining how a universe like ours came to be is currently valued at the odds of one in $\exp \left(10^{124}\right)$. This is the most up-to-date estimate regarding the price to be paid at the Big Bang, if we want to explain how a world as unlikely as ours came to be. 

\subsection{What is our ambition for cosmology?}

Are we done counting states in the Universe's Hilbert space, now that we have, with the last step, added dark energy? This is where our project starts. 

Imagine we are in a future far, far away, say 2000 years from now. At such a time what would we like the achievements of cosmology to be? What would we like to call the accomplishments of our discipline? Would we like to have explained the nature and origin of non-baryonic dark matter, the existence of a non-zero cosmological constant, and the origin of supermassive black holes? Yes! Would we like to have settled the longstanding question of whether to retain, extend, or to modify general relativity, or alternatively to have found it to be an approximation to something else even more compelling? Sure! Would we like to have a theoretical underpinning to the (percent level measured) value of the baryonic density today, as well as an explanation of the observational foundations of the 6-parameter $\Lambda$CDM model?  Of course! 

Ok, so let us proceed a little further. Could we envisage that, in this time far, far in the future, cosmology could have advanced far enough that it could also explain what in the initial conditions to our Universe led to such a complex and unique life diversity on our planet? Freeman Dyson pioneered the asking of such grand questions more than forty years ago \cite{dyson}.

{\it Today}, cosmology is the scientific study of the large-scale properties of the Universe as a whole \cite{dodelson}. We perform an inventory of the contents of the Universe, and study the formation and evolution of galaxies--modelled via point particles in an initially almost homogenous and isotropic fluid with order $10^{-5}$ anisotropies--in order to discern the influence of our cosmological assumptions upon them. Today's probes extend to ever larger volumes, and reach ever smaller-scale inhomogeneities, seeking to constrain and test the $\Lambda$CDM model ever more strongly \cite{dodelson}.

At the same time, as Carl Sagan stated in the very first sentence of `Cosmos' \cite{Sagan}: 
\begin{center}
\textit{``The Cosmos is all that is, or ever was, or ever will be.''} 
\end{center}
Can we imagine that cosmology will one day address {\it everything} we see, including our planet, including our biosphere, including ourselves?

What is our ambition for cosmology? What do we aspire to be (some of) its ultimate goals? Which parts of the Universe do we want to account for, explain, and predict at the Big Bang? What are the seeds that we want to include in the initial conditions, in order to be able to trace their evolution and derive the Universe we inhabit today? 

Can we hope to explain, when we think of initial conditions, why we observe organic structures on our planet that are based on such-and-such a set of elements in the periodic table, and not another, why we observe such-and-such a phylogenetic biodiversity, and why we observe this biome and not a different one, with entirely distinct evolutionary and ancestral origins?

At the same time as we seek the Universe's quantum gravitational origins --- and the corresponding set of initial conditions that has riddled us for so long --- can we strive to understand, in the long run, why the diversity of life on our planet is so unimaginably vast? What was there in the initial conditions that allowed for a planet bustling with such diversity, thriving so creatively with life, forever kindled by creativity and innovation? 

Here, in line with that ambition, we ask whether there is a number we might be overlooking in our inventory of states in today's Universe. Our counting of microstates hitherto, which includes all known fundamental interactions in Nature, looks like a barren moon; the Universe's Hilbert space in our cosmology does not include any microstate with living systems. None of these universes looks remotely like the vibrant Earth we inhabit. 

However given that the entropic content of a system is also a measure of its complexity --- and given that life in the biosphere is unquestionably complex --- we must ask what is the size of the Hilbert space for life? So our question arises, how would the Hilbert space for a living system look if we knew how to compute its microstates from first principles? How many of these microstates, $N_{\rm Bio}$, does the current macrostate of the biosphere contain? Do we have any methodology in physics that is appropriate to compute this number? And if we could compute it, how would it weigh against the remainder of the entropy in the Universe? Can we take the very first steps towards inclusion of biology in our discipline?

\section{Counting the biological Universe} 

The purpose of this paper is to ask the question --- which at first seems to be readily answered with a resounding `no!' --- of whether the `information' contained in life (when measured in terms of the volume of the ensemble of microstates in the biosphere) could make a sizeable contribution to the information content of the Universe, which already vastly exceeds that of its thermal distributions of photons and neutrinos. Could it even approach, or perhaps even saturate, the $S_\Lambda$ bound?\footnote{Here and throughout we refer to {\it `information'} in a loose sense as a quantity that reflects the number of microstates in the biosphere, $N_{\rm Bio}$. We do not attempt to provide a formal definition of information measure in biological systems, like those we define in physics.}

At first sight, demanding the inclusion of life appears obviously absurd. There are far fewer particles on Earth, by some thirty orders of magnitude, than there are in the observable Universe. Even if we take every particle in a classical configuration space, and consider permutations of every particle in the biosphere up to, say, a 10 mile radius above sea level, that configuration space will be minuscule compared to that from $S_{\Lambda}$.  Even if life proliferates throughout the Universe, the particles making it up can be only a tiny fraction of the total particles available and hence their contribution to the `entropy' appears to be utterly subdominant. Hence it seems straightforward that any biological contribution must be negligible. 

\subsection{Is biology physical? Deriving the biosphere}

How then could our biosphere contribute significantly to the cosmic information budget? By the above argument, if one were a committed reductionist, the contribution is negligibly small, necessarily. Reductionism in science can be thought of as the principle that one can best explain something by breaking it further and further down into its individual parts. According to such a viewpoint we could explain all complex phenomena of life on our planet and its entire evolution starting from the onset of life 3.7 billion years ago, by integration of the equations of motion of every individual elementary particle in Hilbert space.\footnote{This is ultimately still limited by quantum uncertainty which inserts genuine randomness into the evolution that may propagate to larger scales by various mechanisms (for instance the radiation-induced cancer of a major world leader, or use of a quantum random number generator to decide which city is to host a new airport).}  That is to say, in an ideal world, if we wanted to include life on our planet in the initial seeds to the Universe, we would start by writing down the equations of motion for the fundamental particles of the Standard Model at the Big Bang, $t=0$. We would then feed these equations into our computers and find that, after 9 billion years, a little rocky planet, the third from the Sun, has developed a primeval slime from which all sorts of ever more complex organic compounds will arise. We would have derived the biosphere and the never-ending complexity of living organisms on our planet. We know that this is not practically possible, for even if we believe reductionist methods would grant us this, we do not have access to such a computer. 

This may at first sight seem a merely technical issue, the only obstacle to the integration of the above equations of motion being computational performance and capacity. This is a practical obstacle, but we think the issue may lie deeper. Indeed in the remainder of this article we want to argue that the endeavour of deriving today's biosphere from first principles, i.e.\ using physics methodologies like the action functional or Newtonian integration of the equations of motion, is simply impossible both {\it for all practical purposes} and {\it in principle}, for several distinct reasons that we will list in Section~\ref{outsideNewtonian}.

\subsection{Condensed matter physics and biology}

Before addressing the issue of how to derive biosphere properties from first principles, let us remark that we do not even have to step outside of physics to hear similar views on the limits of pure reductionist methods. Condensed-matter theorists --- including Nobel Laureates Anderson, Laughlin, and Leggett --- have long contended that in the physics of many-body systems there emerge complex phenomena, and that these are subject to novel {\it emergent} laws. They defend that these laws are every bit as fundamental as those governing elementary particles in that they cannot be understood in terms of any simpler constituents of the system \cite{anderson,leggett,laughlin}. Indeed phenomena such as superconductivity simply disappear as the size of the system is reduced. These views are echoed throughout studies of complex systems, for example Refs.~\cite{Parisiem,Drosselem,Ellisem}. Musser \cite{musser} provides an accessible non-technical introduction to the topic.

As we will argue in Section~\ref{outsideNewtonian}, such emergence of structure is  accompanied by an expansion of configuration space by the addition of the new states that have emerged. This change in volume of configuration space violates the assumption required for use of the Newtonian paradigm --- the stability of the size of configuration space --- as we discuss next in Section~\ref{newtonianparadigm}. In line with that comes the need for a new methodology that accommodates the {\it evolution} of the volume of all possible microstates.\footnote{Debates about the size of configuration space, the number of microstates, and the multiplicity of the corresponding macrostate are mostly of abstract nature. The details of a particular microstate, or of their number, are rarely observable quantities. A macrostate can be experimentally measured and identified, with the validity of this measurement lasting for a small but finite extent in time \cite{wikiBoltz}.}

If such an issue of expansion of phase space arises already at the complexity level of condensed matter, we can expect it to be all the more prevalent as we approach the flourishing complexity of even the simplest living organisms. It would seem that in order to make meaningful further progress as physicists in our understanding of complexity in Nature, we need to extend our toolkit in a way that allows us  to study complexity from a fundamental, first-principles, point of view. 

\subsection{The Newtonian paradigm}   
\label{newtonianparadigm}

In order to proceed with our goal of counting physical states in living systems, let's review the key assumptions that the above calculation of the Universe's entropy relies on: the deterministic and reductionist viewpoints.

The reductionist view has prevailed in fundamental physics for perhaps three hundred years and is valid for systems where the configuration space stays prestated and {\it fixed} and the laws remain {\it unchanged}. For cases where the configuration space changes over time --- i.e.\ states get added on to or subtracted from it --- Boltzmann's entropy formula, which connects logarithmically to the number of possible configurations of the system, ceases to hold. With a prestated and fixed configuration space and the assumptions below, the future evolution of a system is fully determined by the positions and momenta of the constituent particles, via the integration of their equations of motion as determined by the accompanying laws. In summary, the set of assumptions that our calculation of the Universe's entropy above relies on are:
\begin{itemize}
\item The laws are (quantum) deterministic;
\item The dynamical laws are expressed by distinguishing sets of trajectories in configuration space;  
\item The dynamical laws governing the system remain {\it unchanged} throughout the entire evolution of the Universe, and lastly,
\item There is a {\it fixed} configuration space that the states move within, which {\it does not itself change during the evolution of the Universe}.  
\end{itemize}

This set of assumptions forms {\it the Newtonian Paradigm}  \cite{robertolee,lee},  and adds up to the metaphysical hypothesis that there is a single mathematical object, which is given as a solution to a set of fundamental, microscopic laws, such that every property of the Universe's evolution is mirrored by a theorem concerning that object.

Particularly relevant to us, the last assumption that \textit{ ``the configuration space has stayed constant throughout the evolution of the Universe''} cannot be emphasised enough; it will be the breakdown of its validity that will lead us to the results of this article.

\subsection{Methodological preliminaries}
\label{methodprelim}

The introduction of a new methodology for counting biological `degrees of freedom' raises some general questions, some novel, others longstanding, about issues such as the role of {\it emergent} versus {\it fundamental} degrees of freedom, which we address in detail below in Section~\ref{emergfund}. To avoid some common misunderstandings,  we will deal with these issues up front in this section.  Note though that the main results of this and the following papers actually do not depend on the views one may hold about such questions and on reductionism. Indeed, the four present authors hold a range of views on these very questions.

To set the context we make a number of comments here.  These are expanded upon in our companion paper \cite{paper2}.
\begin{itemize}
\item For the study of complexity in biological systems, we regard it as unproblematic that there exists a fixed set of physical laws, which are summarized by the Standard Model of particle physics, quantum mechanics, and general relativity.  These correspond to Feynman's famously expressed ``jigglings and wigglings of atoms'' \cite{Feynwiggle}.
\item In biological systems change occurs for a variety of reasons, over a wide range of timescales. Much of this is settled science, explained by the principles of the `modern synthesis' involving natural selection, genetics etc.\ \cite{modsyn}.  
\item This settled science in biology includes functional explanations, involving specific emergent properties, which cannot all be directly derived from the equations of the Standard Model. Therefore, in practice progress in biology has relied on a mixed methodology uniting reductionist and top-down or functional styles of explanation \cite{Ellisem}. For example, the explanation for why such systems as hearts exist in Nature must include the function that hearts play of circulating the blood, which in turn increases the fitness level of the animal containing the heart and explains why hearts are ubiquitous. 
\item We cannot get to that insight of explaining the existence of hearts by blindly evolving the equations of motion of their constituent atoms. At the same time, the detailed physical properties of a heart are consequences of the laws of physics. The necessity of supplementing reductionist with functional explanations is unproblematic and part of settled science.  Many biologists use reasoning combining a mix of functional, evolutionary, and physical modes of explanation.
\item There are open questions, some related to the role of mechanisms of self-organization, in the construction of options for natural selection to work on, others relating to the origin of life as well as major transitions, such as the origins of prokaryotes and then eukaryotes, and on to many-celled creatures, etc.  
\item The above considerations hold not only in biology but also in physics fields of study, which have their own regimes of complex and non-linear phenomena, where emergent degrees of freedom dominate and na\"{\i}ve reductionist principles may fail.  This was a subject of debate for many years, but recent developments in quantum materials \cite{nature,laughlin} make it hard to deny a role to emergent degrees of freedom. At the same time, it is difficult to regard the Standard Model of particle physics itself as other than an effective field theory.
\end{itemize} 

\section{Why biology must lie outside the Newtonian paradigm}
\label{outsideNewtonian}

Now we come to a key step in our argument, which is that biology cannot be described within the Newtonian paradigm. We identify two distinct reasons for this. The first reason stems from the fact that there is no definition of a configuration space for biology that is stable in time. The second reason comes from the fact that there is no fixed set of laws governing the evolution of biological systems. Indeed, as recent developments in theoretical biology argue, such laws {\it do not exist}~\cite{LMK,eros}.

We address these arguments separately in the next two subsections.

\subsection{Ever-expanding configuration space}

Here we discuss how the useful conceptions of a configuration space for biology are continually evolving in time.

As an example, let us consider  the quantum-mechanical state space of, say, all the biomolecules in a single eukaryotic cell, an amoeba, including all their physical degrees of freedom.  According to quantum mechanics, there must `exist', in some Platonic world of mathematical objects, an exact, complete Hilbert space that contains, amongst everything else, every folding of every molecular bound state that can be made with the atoms contained in the cell. A radical reductionist roots the argument with this assertion, because they believe this Hilbert space ${\cal H}_{\rm Euk, exact}$ exists.   Of course, unlike typical Hilbert spaces we use in quantum physics. we do not know how to construct this space or compute in it. To begin with,  it is clearly of overwhelming complexity, almost none of which is relevant for understanding the properties and behaviour of the actual organism. The problem is that having stated the physical state space, ${\cal H}_{\rm Euk, exact}$, almost none of the states within it have the extremely rare complexity and correlations that come of being alive.

We may try to see if it would help to make a coarse-grained description, still using as observables the operators for the elementary particles.  This would give us a division of the volume of the cell into a large number of boxes, in each of which is specified the density of the various atoms.   The trouble is that once the coarse-graining scale is above the size of typical molecules, you cannot deduce much from the densities of the workings of the cell.  A dead cell and a live cell will have pretty much the same coarse-grained distribution of atoms. To distinguish the molecules of a living cell from those of one that just died will require a microscopic description.

How do we construct a state or configuration space which will allow us to employ our physics tools to reliably predict how a biological system advances in time?  One idea which has proven useful begins with the present state, which is by definition chosen to be viable.  We can then explore a subset of states that differ by a few alterations from the present state.  This is called the adjacent possible \cite{TAP} and it will be addressed in some detail in Section~\ref{tapequation}. Since it depends on the present state, it evolves in time.

The state space then evolves following the evolution of the species.  In some circumstances, natural selection will tend to narrow the number of viable alleles as the less fit are winnowed out.  The result will be a shrinking of the adjacent possible.

Other mechanisms increase the diversity of the viable species; these include the invention of new niches or the branching of a niche into several, exaptation, and speciation. The result is that the diversity of the biosphere, which is a proxy for its configuration space, {\it increases over time}.

But that is not all. We turn now to what it means to write down a law to reliably predict the evolution of a space of (living) configurations.

\subsection{No governing laws at work}

We will argue in this section that it is not possible to reliably predict the next step in the evolution of a biological system: that the biological constituents permanently combine and create novel structures that give the constituents selective advantage, in a combinatorial process.  Note the exquisite nature of the principle of `selective advantage' which is at play in biological networks and governs organization within the system. Only extremely rarely is an equivalent of selective advantage invoked in physics as a environmental controlling factor; examples are Ref.~\cite{LeeBH} in which one of us (L.S.) applies the principle of natural selection to the birth of universes by positing that the collapse of black holes could lead to the creation of a new universe with slightly different values for the fundamental constants, and the self-reproducing inflationary Universe \cite{LindeSR}.

For example, a fish develops a swim bladder for the first time, and there happens to be a bacterium that floats by and realises ``that's a great place to live in''. Suddenly the bacterium is selectively favoured due to the newly-found environment and has created a novel bound state with the fish. That the bacterium happened to swim by and found a hospitable environment in the fish's bladder could not have been predicted before, nor written down {\it a priori} in the form of a law.

We have found that in biological sciences it is a true challenge to write down underlying governing laws that can stand in for a {\it standard model}, in the fashion we look for in theoretical physics. Meaning laws which, whether in deterministic or probabilistic form, can be turned into equations of motion for predicting biological complexity. We have exhaustively debated this search for an underlying principle or law that could stand in for a fundamental rule in biology from which we can predict the system's next state.\footnote{That is to say over significant representative time and size scales. Naturally, if we sufficiently isolate a system both in size and time we will recover the behaviour dictated by standard laws of the relevant physical theories.}

Ultimately we could find only one rule that applies and underlies behaviour and evolution in {\it all} living systems without exception:
\begin{center}
 {\it The name of the game is getting to exist.} 
\end{center}
We have found that any deviation from a rule as generic as this promptly fails to work in one or another set of circumstances. We can state no principle more specific than this one, that applies globally as an all-encompassing rule for biological evolution.

Since there are appear to exist no fixed laws \cite{eros}, it turns out that the concept of a biological law is then necessarily very different from that of a law of physics. It must be able to characterize the evolution on a trajectory in a configuration space {\it that is itself dynamically evolving}. We cannot hope to answer questions in biology by integrating the equations of motion. For the reasons we argued here, pure physics is not sufficient to explain complex organization by itself.

We don't wish to be dismissive about the considerable challenges that the irregular characteristics of biological systems we have described pose to our understanding of them within physics.  Our goal here is to take the first steps by introducing tools into our discipline --- namely combinatorial innovation and the TAP equation --- that will allow us to address systems that exhibit unprecedentedly high degrees of complexity and self-organisation. Our aim is that, by incorporating these and related tools, our discipline will evolve to encompass in its field of study these highly non-ergodic, highly-complex systems and gain an understanding of them based on a physics, first-principles, approach.

\section{The role of effective theories in physics and biology}
\label{s:EFTs}

In this section we ask whether we can approach the counting of microstates in biology in the same fashion we approach effective regimes and effective field theories in physics, which prevail within the strict limits of their regime of validity. The role of effective theories from a biological perspective has already been discussed in Ref.~\cite{Ellisem}; see also Ref.~\cite{Noble}.

Our understanding of Nature in physics is based on the assumption that there exist key postulates and principles at work in physical systems, and our aim is to derive the (effective or fundamental) laws governing the system in order to predict its future behaviour. This is the Newtonian paradigm at work. 

We then investigate whether we could search for an underlying fundamental theory of biology by using the effective field theory approach we use in physics.

\subsection{Effective field theories in the Standard Model of particle physics}

When we come to count the number of possible states in biology it will be helpful to recall the analogous situation of counting states in effective field theory in physics. This example will be crucial in identifying clearly the difference between the evolutions of physical and biological states.

When asked to count the number of possible microstates in the Universe, compatible with the one current macrostate of today's, the first item to settle upon is what interactions are to be including in our description of the Universe. The total number of possible microstates and consequently its entropy depends on the number and nature of interactions which are accounted for. 

Further yet, the number of states for the Universe obtained will depend on the regime the Universe is in, whether it is effective or fundamental (meaning we have a choice on how to treat the Hilbert space relevant for cosmology at a given time). Let us give an example of the kind of emergence of different regimes in biology by comparing it with a similar phenomenon in physics. 

In cosmology the relevant Hilbert space gives rise to different effective field theories:\footnote{Cosmology is the study of the evolution of the Universe through all the effective field theories of the Standard Model, from one to the next, to the next. Therefore it cannot be argued that we need only take one subset of states at a time. Because cosmology is a history, all subsets of the Hilbert space, at all temperatures, need to be present.}
\begin{itemize}
\item If we take an Effective Field Theory (EFT) point of view, in the early universe, at high temperatures, bound states are absent or rare. We can then to a good approximation find the entropy by counting just free particle states, without worrying about the effects of the electromagnetic charges. This is EFT1.
\item When the temperature falls to that of cosmic nucleosynthesis, $T_{\rm nucleo}$, we need a different EFT that can describe the nuclear bound states or at least the most bound nuclei.  This is EFT2.  EFT2 has more states than EFT1,  because it has discrete series of bound states (deuterium, helium-3, helium-4, lithium nuclei etc.), and these are represented by discrete line additions to the energy spectrum. 
\item At some still lower temperature, more bound states open up, which are atoms, thus the Hilbert space grows by the addition of discrete series of lines for all the atomic spectra.  This is EFT3.
\item And so on \ldots
\end{itemize}
Our point here is to highlight that when new regimes are entered, be they effective or fundamental, new effective interactions emerge. Along with the new interactions comes the emergence of new {\it bound} states which hadn't previously existed. Therefore new states are added to the configuration space, which grows in size. We go back and revise the previous number, our estimate of which was based on incomplete knowledge of the behaviour of the system when new bound states open up. The attributed value for the system's entropy gets thereby updated.  

So far this counting of states can be applied to both physics and biology but now comes a detail of vital importance, which is the key to the argument of this work. When counting states in physics, we have access to the underlying completion of the theory, i.e.\ the Standard Model of particle physics, or its suitable extension, from which the effective laws of the different low-temperature regimes can be derived and predicted.\footnote{We take here, for the present purpose of comparison with methodologies in biological sciences, the view that the various EFTs that the Universe goes through can be derived from the Standard Model (SM) of particle physics in its present shape or ameliorated, and setting aside the various known issues of non-unitarity or asymptotic freedom. In reality several leading contemporary particle theorists no longer believe in the existence of a final fixed fundamental  QFT, from which all the EFT's are derived in the style envisaged by Weinberg \cite{Weinberg}. Firstly we can argue that the SM itself is not unitary at all scales because of the existence of Landau ghosts in the spectrum. Secondly, its couplings are not asymptotically free. The usual Grand Unified Theories such as SU(5), SO(10) are also not asymptotically free, and the supersymmetric SM does not offer much help. Many particle theorists believe that, given its internal inconsistencies, the SM in its exact form is itself also an EFT. In these terms we would be left with three views: 1)  All realistic QFT's are EFT's, and hence there is no complete QFT which completes the SM.  2) Some complete unification, is discovered, e.g.\ what string theory is hoped to be. 3)  Laws evolve in time, as argued in Refs.~\cite{LeeBH,LeeBHbook}. However for the current purposes of establishing an analogy that we can export to biology, the parallel with a common, underlying completion of the SM of which all the different EFT's we may use are approximations, is sufficient to be used in our argument.}

The above EFTi are all different regimes and effective theories. However, and most importantly, the claim in standard physics is that all physical regimes are derived from a single complete microscopic theory, which has a single Hilbert space that does not evolve in time or depend on temperature. This Hilbert space, $H_{\rm SM}$, contains all the free and all the bound states, according to the completeness relation, and doesn't evolve. Neither does the dynamics. It is this non-evolution of the Hilbert space that allows us in cosmology, as we argued in Section~\ref{newtonianparadigm}, to recover the value of the Universe's entropy, together with the measure of the set of initial conditions.

\subsection{The missing standard model of biology}

The same will not hold for biology. It is the case that a similar situation occurs, with new bound states emerging spontaneously from simpler constituents and with them new effective interactions.  And so, similarly, when a new effective biological regime emerges ---  for instance when a group of organic molecules assembles, in a detailed combination, to give rise to a cell --- we have entered a new regime of the system  \cite{Kbeyond}. For now the processes governing the behaviour and evolution of the cell, heritable variation and natural selection, are no longer those governing the behaviour of organic molecules. Nor will we learn anything about the evolution of the cell by examining the equations of motion of a composite of $N$ organic molecules comprising the cell.

In order to understand this, let us then examine what could be analogous within our own biosphere to the sequence of EFT1, EFT2, EFT3, \ldots
\begin{itemize}
\item complex organic molecules
\item amino acids
\item proteins
\item organelles
\item proto-cells
\item cells
\item eukaryotic cells
\item multi-cellular organisms
\item \ldots
\end{itemize}
In each of these regimes a new bound state has emerged in the form of a novel structure, made up of free states of the previous regime. It is an emergent bound state, just like the emergent bound states in EFTi of the particle physics Standard Model, and it carries with it a new effective interaction that will govern evolution in the new regime. 

And here comes the difference: while in physics we have (or can imagine having) a completion of the Standard Model of particle physics from which to derive different effective regimes and laws, in biology we do not. There is no underlying model echoing the Standard Model of particle physics to determine or predict the new emergent interactions from the simpler regime. We cannot derive the behaviour of a complex group of living cells from a set of fundamental rules. Moreover, life in other biospheres may follow completely different paths.

The reason is that the configuration space, the number of possible states in a biological system, is not stable like in the Newtonian paradigm in physics. The number of possibilities, the number of possible new states, is always increasing in many available and {\it unpredictable} directions \cite{kauffschrod}. The (biological) Universe is highly non-ergodic and so to fully explore the configuration space of the biosphere would need, even at a Planck time per state, impossibly many Hubble times \cite{paper2}.

If the biological configuration space cannot be controlled and is always growing, we cannot predict or even deduce the next step in the evolution of the system. One of us (S.A.K.) has gone even further beyond this statement, and argued (with A.~Roli) that such an underlying model of biology will never exist \cite{eros}. The claim is that the effective regimes that emerge in biology will never be derived from fundamental (biological) principles. 

That observation leads us to formulate the most important claim in the current work,
\begin{quote}
\textit{If new effective interactions, and their associated novel bound states, are continuously emerging, as is the case for living systems, we are never finished counting possible states, as their associated Hilbert space is ever growing.}
\end{quote}
There lies, in summary, the bulk of our argument. Such developments in theoretical biology indicate that the number of possible new emergent interactions and associated emergent states in biology grows continually as a secular trend, and is therefore unbounded. This implies there does not exist, even in principle, a standard, microscopically complete, model of biology. Or, in alternative, it exists but will never be completed. We do not have any fixed standard model of biology to predict what the next level of emergent organizational complexity will be. In biology we are but observers of the dynamics, in puzzlement or wonder. 

\section{Emergent versus fundamental degrees of freedom in physics}
\label{emergfund}

In the last section we concluded that, due to the lack of an underlying law of biology to derive them from, we cannot use the technique of effective field theories in biology. But the fact remains that there are `degrees of freedom' which emerge in each new regime of biological systems. These `degrees of freedom' constitute the key variables relevant to description of that regime. Given that we established in the previous section that these emergent `degrees of freedom' cannot be derived from a fundamental theory, the question we address here is whether it is fair to say that those emergent `degrees of freedom' should then be regarded as fundamental variables in themselves, corresponding to so-called {\bf strong emergence} \cite{Drosselem,Ellisem}.

Our goal in this paper is to count the number of physical microstates of the observable Universe, which are the ones contributing to its `entropy'. We reviewed the contributions to entropy from cosmological components that are well understood and plausibly fundamental, reaching the conclusion there is roughly $\exp (10^{90})$ in the free particles, plus $\exp (10^{104})$ from counting those degrees of freedom whose values are trapped in black holes, and finally $\exp (10^{124})$ for those trapped in the cosmic horizon, $S_{\Lambda}$. 

We then asked whether biology, with its organizational complexity, could supply additional emergent `degrees of freedom', and thereby contribute extra microstates, $N_{\rm Bio}$, to configuration space. We asked whether that contribution could be numerically large enough to perhaps  dominate the known ensembles of the Universe's Hilbert space, or even enough to approach or saturate the $N_\Lambda$ bound.

As we argued above this seems absurd at first.  To consider this as a possibility one has to understand Nature while regarding fundamental and emergent `degrees of freedom' as equivalent.

We believe that the evidence is strong that this is what we must do.

The great tool that physics has to contribute to the study of macroscopic systems is the reasoning, and point of view, of statistical physics.  This of course leads from the microscopic description of a physical system through to deductions of its bulk properties, through a device which is the statistical ensemble.   It is very clear from the studies of many quantum systems that the quantum and classical descriptions of emergent `degrees of freedom' have to be counted as equivalent to fundamental degrees of freedom, from the point of view of their thermodynamics \cite{anderson,leggett,laughlin}.

Before detailing the arguments it is important to emphasize that the claims we are about to make do not depend on what one believes about the primacy of reductional versus functional and other types of explanation  in quantum field theory.  The fact that the state of a biosphere or a living organism is radically anti-ergodic, in the sense that there is an  enormous degeneracy near the ground state, arising from different utilization of proteins evolving to be used for different functions, does not depend on the philosophy of the scientist. It is also a fact that almost none of these states are occupied in any given realization.

A reductionist will describe this in terms of a universal, time-independent, Hamiltonian, with all these states in its spectrum. But they will no more be able to compute that spectrum in advance than will their effective field theoretic cousin.  Both will have to refer to partly functional explanations, because those are the best available explanations, even with the limitation of being not deducible (though they disagree about why that is). One will refer to newly occupied states,  the other  to newly emergent states, but each will do the same calculations in support of the same explanations.

In the rest of this work we will argue that composite or collective `degrees of freedom' could in some circumstances, particularly in biology, greatly outnumber fundamental degrees of freedom. 

We have stressed that biological and cultural evolution take place in a region of phase/state space which is {\it highly non-ergodic}, in the sense that the distribution over different molecules is highly sparse, irregular and complex. We discuss this extensively in our companion article, Ref.~\cite{paper2}.

Consider for example a part of state space that consists of all bound chains of 200 or fewer amino acids in the biosphere. Life on Earth uses about 20 different amino acids, so there are roughly $20^{200}$ states in its configuration space. Almost all are unoccupied because they could not form by random processes in many times the history of the Universe.  This is what we mean by biological evolution being highly non-ergodic in its configuration space. The few states that are occupied are those that are produced by long, specific sequences of catalyzed reactions, and those are ones which grant some form of functional evolutionary advantage to the organism that the amino acid chain is a part of. 

Consider that any classical physical object has diverse causal properties, each of which can play a functional role.  The same protein can evolve to be used in a cell to catalzye a reaction, to bind a ligand, to absorb a photon, to carry a compression or tension load, or to be a strut upon which a molecular motor can walk. Indeed in evolution transitions between these functions arise all the time.

It is critical to our discussion that any given protein existing in the biosphere is, as a physical entity, a perfectly sensible set of physical degrees of freedom, such as the positions and momenta of all its atoms. However, that a specific protein `gets to exist' in the biosphere is because one or more of its causal features were selected to play one or more functional roles. Thus we can sensibly count the physical degrees of freedom in the space of all proteins up to amino acids. But we cannot count the number of possible functions \cite{eros}. Thus we can define the physical `entropy' of the biosphere but we cannot define its functional `entropy'. Yet it is precisely the functions of proteins in cells that, together with physics, accounts for the existence of those proteins in the Universe.

Only once one knows that a particular protein chain is represented in configuration space can we {\it verify} the story of how it emerged and was selected and sustained by heritable variation and natural selection or genetic drift, which at every step will be consistent with the fundamental laws of microphysics. But one could never have deduced that exactly that particular novel chain would emerge and that that function would be carried out by that particular polypeptide (which most likely is far from the optimal polypeptide to provide that specific evolutionary advantage).

What we are saying is that one could build a wavefunction or density matrix for the regime of Hilbert space which contains all sequences up to 200 amino acids, and one could write down the operator that measures the density or occupation of the different species. Nonetheless, one is not going to be able to compute the exact distribution function ${\rm Tr}[ \rho H]$ for $1,000$ years into the future. There is a vast sensitivity to initial conditions, and there are any number of events that strongly influence which small proportion of possible proteins or possible animals, get to survive for a time around our time. One may believe that this is possible in principle, but it is evident that no computer that could fit inside the observable universe, running for the Hubble time, could come close. There are also issues of whether round-off errors from digital approximations to fixed numbers of decimal places converge \cite{zwart}.

The fact that the occupation numbers are distributed in this sparse and complex sense confirms that we are also in a regime which is far from thermal equilibrium. This reminds us of the many works, e.g.\ Ref.~\cite{Morowitz}, which argue that the biosphere has its origins and flourishing due to a slow, steady flow of energy through the biosphere.

What we are arguing here is that the biological configuration space generated by the emergent functional `degrees of freedom' vastly dominates in number the space generated by the fundamental degrees of freedom. This is particularly true in the circumstances of exquisite non-ergodicity, as well as exceptional organizational complexity where life is possible. 

The foregoing set of arguments has been suggested by one of us (S.A.K.) and collaborators, as we mentioned already in Section~\ref{s:EFTs}, as leading to the conclusion that no entailing laws govern the evolution of the biosphere \cite{LMK,eros}. This might seem too strong a conclusion, because it implies that we can make no prediction for the number of states in the biosphere that emergent `degrees of freedom' can create; that number may simply be indefinite.

However, as physicists we have a lot of experience turning indefinite numbers into definite finite numbers by imposing constraints which act as regulators that express the quantity in question as the limit of a sequence of finite well-defined numbers with physical meaning. In the subsequent sections we carry out such an analysis to produce a lower limit on the number of different microstates, $N_{\rm Bio}$, potentially existing in the biosphere consistent with the biosphere's macrostate today. This number will give us an estimate of the number of microstates in the configuration space of life on the planet, and produce a lower limit to $N_{\rm Bio} (t)$ by limiting the counting to those states or structures which are produced by iterating or otherwise combining existing states, through combinatorial innovation, in the process described by the Theory of the Adjacent Possible, addressed in Section~\ref{tapequation}.  

We looked here at examples where the number of emergent degrees of freedom vastly exceeds that of the fundamental degrees of freedom. In such cases it may not be possible to excite simultaneously all the physical degrees of freedom, due to a shortage of materials (as well as shortage of  cosmological time). Such systems cannot be in equilibrium and they cannot be ergodic. We address this in fine detail in Ref.~\cite{paper2}. Nonetheless, in biology just like in physics, we count all the possible excitations of the ground state as different states in the statistical ensemble that the real system might have been in.

\section{Permutational microstates in biological hierarchical models}

In this section we investigate whether we can find systems in biology or otherwise for which the counting of new microstates (obtained through {\it permutation} of its constituents elements, as is practice in the Hamiltonian or Lagrangian formulations of mechanics) can exceed the magnitude of the number of states corresponding to the holographic bound. As an example in classical mechanics the parameters used to identify a particular microstate whose constituents are point particles are the values of its $6N$-dimensional generalized coordinates. 

So the next step is to investigate how the `information' encoded in the arrangements of biological molecules (their generalized coordinates in this case) could exceed their thermal `entropy', if molecules in living systems were disassembled into their component atoms and allowed to come to thermal equilibrium with the cosmic microwave background (barring out the entropy of gravity and of other interactions that we know play a role at the cosmological level).

We begin with two examples that illustrate that hierarchical systems can, through organisation, develop vast numbers of possible states, which may however be subject to material limitations on the ability to construct them. However, we will find that permutational operations alone do {\it not} suffice to give a route to exceed the holographic bound. Indeed we will conclude at the end of this chapter that only the unprecedented, hyper-explosive growth in numbers achieved through combinatorial innovation, which will be introduced in Section~\ref{tapequation}, can come into view as a worthy competitor to the magnitude of numbers in the volume of $\Lambda$'s configuration space. 

\subsection{Three thought experiments}

\subsubsection{Thought experiment A: a simplified cell model}

Let's start with a basic example: a simplified model of a biological cell.\footnote{Here are some useful comparative numbers for our actual biosphere. Over 99\% of species that ever lived are believed extinct. There are estimated to be 10 million species on Earth, mostly undocumented. The most prolific living organisms are a type of marine bacteria of which there are about $10^{28}$ individuals. There are about $10^{10}$ humans, a total of $10^{11}$ birds (mostly domestic chickens), $10^{13}$ fish, $10^{19}$ insects (including $10^{17}$ ants), and $10^{12}$ trees. There are $10^{31}$ phages (a virus that infects bacteria and archaea) but these are not living (at least, cannot reproduce on their own). A human body contains about $10^{13}$ cells.} For present purposes we simplify the cell down to the system of DNA, RNA, and proteins, although of course many of the proteins function to catalyze chemical reactions that do the other work of the cell.  We posit that our cell has $10^3$ genes, with each gene made up of $10^3$ triplets coding for the amino acids. So the genome is $10^6$ amino acids long, and there are then
\f
N_{\rm genomes} = 20^{10^6}  \simeq 10^{10^6}
\ff
possible genomes for a cell. Since this number already considerably exceeds the number of atoms in the observable Universe, the set of genomes is a highly non-ergodic system where only a tiny subset of possibilities can be actualized \cite{paper2}. Additionally only elements of the sparse subset that corresponds to viable self-reproducing organisms can be expected.

For this first example we assume that life never makes it beyond single-celled creatures, and also ignore gene expression so that the genome completely determines the cell. We might imagine that this corresponds to the first two billion years of the development of life on Earth.

Even then, to construct a biosphere we have $N_{\rm genomes}$ possible species available. A subset of these make up the actual species present; there are
\f
\label{e:allbiouniverses}
N_{\rm imaginable\;biospheres}= 2^{N_{\rm genomes}} \simeq 10^{10^{10^6}}
\ff
ways of selecting these subsets. 

This impressively-large set however takes us far beyond even non-ergodicity! If we select a single typical item from this set, it contains far too many species to be made from all the material in our entire observable Universe. Certainly, we wouldn't want our biosphere to be so densely populated as to collapse into a black hole! In other words, such heavily-populated biospheres are not valid physical states. (This is consistent with Banks' second argument for the holographic bound, which envisages attempts to build high-entropy systems being thwarted by their collapse to black holes \cite{banks}.) Hence not only can we not ergodically investigate the entire set, but we cannot even actualize most individual members of the set. These biospheres are simply impossible, unless we modify the large-scale properties of the Universe. We have called them `imaginable' because we can envisage them and their properties, while accepting their impracticability. The existence of cosmological limits --- of which this is only one of a number we will encounter in the course of these papers --- justifies our proposal that biocosmology be considered a new scientific field with a non-trivial subject matter.

To be more realistic, we can use Earth as an example and limit the number of species to $10^7$, guided by current estimates of Earth's biodiversity. Still, each of those species could be any one of our genomic options, yielding the still very substantial estimate
\f
N_{\rm realizable\;biospheres} = N_{\rm genomes}^{10^7} \simeq 10^{10^{13}} \,.
\ff
Of course many ecosystems may prove unviable even if composed of potentially-viable species.

This estimate counts species without regard to how many individuals each species has. Again guided by Earth, single-cell populations upwards of $10^{25}$ per species are certainly reasonable, so a given biosphere of $10^7$ species has at least $(10^{25})^{(10^7)} \simeq 10^{10^8}$ variants with different population distributions. However the exponent in the above equation was already so vast that multiplying by this number doesn't change it!

We learn something vitally interesting from this example, which is that even at this low order of complexity, cosmology puts very strong constraints on biology.   Biology is, in some sense, the exploration of the possible self-organizing entities.  It becomes apparent that this exploration is mainly resource limited, and that the ultimate resource-limiting constraint is the size of the Universe itself. 

We will continue counting different measures of the possible size of  the biological world.  We will find two kinds of numbers. We have those that are {\it comprehensive}, in that they count all the possibilities of structures that might be made at the appropriate level of complexity.   Then we will find  measures of complexity that are limited by the resources available in the Universe, such as the number of atoms or stars.  We will call these {\it resource limited.}

A very interesting pattern characterizes these numbers. The comprehensive counts often vastly exceed the Bekenstein entropy of the cosmic horizon \cite{Bekenstein,Bek2}, given by $N_{\Lambda}  = \exp (10^{124})$, while the resource-limited numbers are dwarfed by it:
\f
N_{\rm resource\;limited}   \ll   N_{\Lambda}  \ll N_{\rm comprehensive}   \,.
\ff

Let us hypothesize that this is correct. Let us also recall that biological systems often grow to dominate their environments, so that what most often limits the exploration of different levels of organized complexity is the resources available.  The implication is that, in the long run, cosmology will be a greatly-limiting factor on biology.

Let's go up further levels of complexity and see if this pattern repeats. At the next  level of complexity, we assemble our observable Universe from subsets of biospheres. These presumably, largely if not completely, evolved independently. For the purpose of argument, let's nevertheless assume they all are based on the same DNA/RNA/protein chemistry.   If all subsets were possible we would obtain
\f
\label{e:compbiouniv}
N_{\rm imaginable\;biouniverses}= 2^{N_{\rm biospheres}}   
= 2^{10^{10^{10^6} }} \,.
\ff
once again shifting our count up an exponential level.  We find ourselves again in the region far above $N_{\Lambda}$. Again we can impose a material limitation, limiting to one biosphere for each of the $10^{22}$ or so planets in the observable Universe, to get
\f
\label{e:sphere2univ}
N_{\rm biouniverses} = N_{\rm biospheres}^{(10^{22})} \,.
\ff
To be consistent we should also restrict to the realizable biospheres to find 
\f
N_{\rm realizable\;biouniverses} = 10^{10^{35}} \,.
\ff
While exceedingly large, this estimate of realizable biouniverses is still considerably within the number of states allowed by Banks's holographic bound, which states that $N_{\Lambda}$ bounds the number of states within the observable Universe \cite{banks,holo,bousso} (though see also Ref.~\cite{Leeholo} for an alternative view):
\f
N_{\rm realizable\;biouniverses} = 10^{10^{35}}  \ll  N_{\Lambda}  \ll 
N_{\rm imaginable\;biouniverses} = 2^{10^{10^{10^6}}} \,,
\ff
even though the comprehensive number of possibilities far exceeds it. 

We shouldn't be too surprised by this, because we haven't invoked any actual properties of living systems or the associated expansion of state space. We thus remain within a combinatorial physics setting, and in limiting ourselves by the number of particles available in our Universe we are within the regime of validity of the arguments leading to the holographic bound. The vast set whose size is given by Eq.~(\ref{e:compbiouniv}) cannot be meaningfully compared with the holographic bound because that bound is a property of our own Universe, within which those elements cannot exist.

We next investigate the microstate numbers that can be obtained by extending this analysis to include multi-cellular organisms with a variety of gene expressions.

\subsubsection{Thought experiment B: multi-cellular biospheres}

Now we consider the genome of a multicellular creature. We go up a level of complexity, in that the organism has a large number of cells,  let's say $10^9$.    These all contain the same genome, but read off  different subsets of the proteins coded for by that genome.   This is possible because there is a genetic regulatory network that turns on and off the different genes.   This creates a new level of the hierarchy, which is cell types.   Each cell type has a distinct subset of the genes activated  (at any one time, but we ignore that complication here).   Thus there are as many potential cell types as there are subsets of the genes of the creature. There are then up to 
\f
N_{\rm cell\;types} = 2^{N_{\rm genes}} = 2^{1000} \approx 10^{300}
\ff
possible cell types, given the original set of genes.\footnote{We are being generous here in allowing all permutations; usual lists of human cell types contain only hundreds of entries. But this number turns out not to be crucial to the end result.} This number is big enough to ensure that every cell and hence every organism, even from the same genome, is unique. Indeed, `identical' twins are never precisely identical. 

The number of possible individual creatures constructed from a single genome is then
\f
N_{\rm creatures} = (10^{300})^{10^9} \simeq 10^{3\times 10^{11}} \,,
\ff
and an assumed population of $10^{10}$ of them can be drawn in
\f
N_{\rm populations} = (10^{3\times 10^{11}})^{10^{10}} \simeq 10^{3 \times 10^{21}}
\ff
different ways. The number of cell types gives only a small modification to the final answer since by construction cell-type diversity, driven by gene number, is much less than genomic diversity which is driven by amino acid number. Hence it could have been ignored.

The construction of a multi-cellular biosphere now starts by associating one of these possible populations to each genome present. Let's imagine $10^6$ multicellular species present, giving $N_{\rm genomes}^{10^6}$ possible species sets.  Then there are 
\f
N_{\rm multicell\;biospheres} = 
\left(N_{\rm genomes}\times N_{\rm populations}\right)^{10^6}
 = (10^{3\times10^{21}})^{10^{6}} \simeq 10^{3 \times 10^{27}}
\ff
realizable multicellular biospheres. In this case the variety of genomes proves irrelevant, being dominated by the population diversity. 

We can again convert to realizable biouniverses using Eq.~(\ref{e:sphere2univ}), getting a number much bigger than the single-cell case but once more well within the holographic bound, in agreement with the pattern we identified in the previous discussion. In all these cases the exponent is essentially given by the total number of cells in the combined system, up to some modest correction factor, hence evidently is resource-limited by the material available to make cells.

\subsubsection{Thought experiment C: Lego Star Wars kits}

We continue with a further thought experiment, which we call {\it the Lego Star Wars experiment.} Even if not a living system we deem this example a good illustrator of the permutational potential for creating vast numbers of novel microstates.

You can't buy such a set today, but if you go into the Adjacent Possible Lego Store you can purchase a 3-in-1 Star Wars Lego kit. It contains perhaps $N=300$ plastic pieces; some the familiar lego blocks, but mixed with a diverse variety of special pieces.  The instructions are in one or more books.   They offer you the steps to construct, with the pieces available, one of several different toys: a starship, a speeder, a walker.   You snap them together and remarkably each toy requires the same set of parts.  

What is interesting is that the same parts can be snapped together to construct several different toys. Indeed any child will quickly get bored recreating the vehicles in the instructions, and will in an hour or two create toys of their own imagination and design. So let's say after a long time spent trying different attempts we conclude that there are $M$ ways the pieces can be snapped together to give a stable composite object.

Imposing an ultra-violet cutoff $a$ (the size of the pieces) and an infra-red cutoff $L$ (the size of the room we are building them in), originally the Hilbert space of the pieces, assuming they are light, is finite dimensional with dimension
\f
d_0 \approx \left ( \frac{L}{a} \right )^{3N} \,.
\label{d0}
\ff
This corresponds to a volume of phase space, $\Gamma$, which is
\f
V_\Gamma = L^{3N} \left ( \frac{\hbar}{a} \right )^{3N} \,,
\ff
measured in units of the Planck constant. This displays the basic rule that if the phase space $\Gamma$ has finite volume $V_\Gamma$, with $6N$ degrees of freedom, the dimension of the corresponding finite-dimensional Hilbert space is
\f
d \approx \frac{V}{\hbar^{6N}} \,.
\ff

The Hamiltonian describes $N$ pieces which have translational and angular momentum,  plus potential energy barriers to their coinciding, which however have in them some channels where they may be snapped together.  A snap corresponds to a barrier of at least $U_{\rm snap}$;  once snapped the pieces are free to move together, with a bound-state energy degenerate with the ground state in which they are all independent. That is, when the pieces are snapped together to make a toy, there are in a state degenerate with the ground state, which we will set at $E_0=0$. The toy when fully snapped together then corresponds to a $d_{\rm toy} = 6$ dimensional region of a configuration space.  The effective low-energy Hilbert space, which consists of states nearly degenerate to the ground state, is the quantization of this sector, and  is also $6$-dimensional.   The rest of the states in the original Hilbert space for the pieces are disconnected by very high energy barriers which you have to get over either by snapping or breaking apart the pieces.

However now let us recall that there are a number, $M$,  of different toys that can be snapped together.   Each excludes the others, so the classical configuration space has $M$ regions which are degenerate with the ground state. Quantum mechanically this implies that the Hilbert space is the product of a gapped subspace and an ungapped  low-energy effective Hilbert space, which consists of wavefunctions that have support entirely in the region degenerate with the ground state.  That is
\f
{\cal H}_{\rm full} = {\cal H}_{\rm deg} \otimes {\cal H}_{\rm gapped} \,,
\ff
where 
\f
{\rm dim}(  {\cal H}_{\rm deg})  = M \times d_{\rm toy} \,,
\ff
and the ungapped sector ${\cal H}_{\rm deg}$  consists of wavefunctions with support on the different toy regions.

Note that the dimension of ${\cal H}_{\rm deg}$ has higher dimension than the naive Hilbert space by a factor of $M$.    This shows that the Hilbert space of possible states can be much larger than the Hilbert space of actual states.

As the cell examples showed us, when we have lots of pieces we can expect the ways to snap them together to be far greater than the number of pieces itself. Now imagine that there are $N=10^8$ pieces and  $M=10^{1000}$  different ways to snap them together to construct a state roughly degenerate with the ground state.  The {\it microscopic Hilbert space} has  dimension roughly
\f
d_0 \approx \left ( \frac{L}{a} \right )^{3 \times 10^{8}} \,.
\ff
But the degenerate ungapped sector of the Hilbert space has dimension 
\f
d_{\rm deg} \approx \left ( \frac{L}{a} \right )^{3 \times 10^{1000}} \,.
\ff
The number of distinguishable subsets goes up really fast. 

\subsection{Principles of hierarchical constructions}

The key thing we have learned is that there are systems which can be organized in a vast number of different arrangements that are all degenerate or near degenerate to the ground state.   Furthermore, the number of ways in which they may be so arranged vastly exceeds the number of parts or basic units.  This means that the degrees of freedom which are inherent in the possibility of being rearranged vastly exceeds the number of degrees of freedom of the parts, considered separately.

What is remarkable is the way that by iterating a process of construction of a hierarchy of complex systems, where the possible entities at the $(N+1)$'th  level of complexity are given by the subsets of the entities at the $N$'th level, one gets the possible numbers of biological systems up to truly astronomical magnitudes.   We will formalize this observation in terms of a class of equations invented in Refs.~\cite{TAPeqn,TAPeqn2} called the TAP equation, studying it here in Section~\ref{tapequation} and in more detail in a separate paper~\cite{TAPmath}.

One way this can happen is if there are a small number of initial parts, $M_0=\{  x_i \} $, $i=1,\ldots M$,  and a rearrangement consists of a selection of a subset of $M_0$.   There are
\f
M_1 = 2^{M_0}
\ff
such subsets.  The choice of a subset is the first level of structure, and it is easy to see that there are more composite objects in $M_1$ than there are objects in the original set $M_0$.     

We can keep going.  A second layer of a hierarchy of construction is to pick subsets of the set of subsets.  This allows for
\f
M_2 = 2^{M_1} = 2^{2^{M_0}} \,.
\ff
There may not be enough original pieces to build many of these composite systems, but that does not prevent these rearrangements from being on the lists of possible structures that might be made if sufficient material could be gathered. At the effective level, where we are interested only in the degenerate sector of the ground state, they are states. 

We note that there is a principle of opacity.  Once we put $a$, $b$, and $c$ in a combination,$A= (a,b,c)$, those parts $a$, $b$, and/or $c$ can be  used again, perhaps in combination with $A$, in a second-level subset of the form $E= \{ a , \{ a, b , c \} \}$. We cannot do a low-energy  experiment that detects the $a$ inside $A$, so that for the purposes of a low-energy EFT that describes the degeneracies of the ground state, they function as independent, unrelated entities, i.e.\ the completeness relation in a quantum effective theory will be
\begin{eqnarray}
I &=&  |a \rangle \langle a |  \;\; \oplus   \;\; |b \rangle \langle b |  \;\; \oplus  \;\;  |c \rangle \langle c |   \;\; \oplus   \;\; 
 |\{ a, b  \}  \rangle \langle \{ a, b \}  | 
 \nonumber  \\
& & \oplus  \;\; |\{ b,c  \}  \rangle \langle \{ b,c \}  |  \;\;  \oplus \;\;   |\{ a, c  \}  \rangle \langle \{ a, c \}|  \;\; \oplus \;\;  
 |\{ a, b, c \}  \rangle \langle \{ a, b, c \}  |  \,.
\end{eqnarray}

The key point is  that a system, which originally has a small number $M_0$ of kinds of degrees of freedom, can be put under the influence of a potential $V_0$ that creates, through a construction that groups the $M_0$ original degrees of freedom into subsets, a larger number of states, $M_1 = 2^{M_0}$ all degenerate with the ground state.  But this does not prevent them from being represented as independent degrees of freedom, as basis states of the Hilbert space that describes the effective theory of the ground-state degeneracy.

Of course these degrees of freedom may not be able to be all excited simultaneously, due to a shortage of materials, and in the more extreme situations even the individual items may not be actualizable within available resources. Thus we are dealing with regimes whose configuration spaces are very sparsely populated, and for that reason are far from ergodicity and equilibrium. This raises the question: why do some biological molecules come to exist in the biosphere, when the vast numbers of others of similar atomic content do not? We develop the implications of this in the following paper of this series \cite{paper2}.

By enumerating all possibilities regardless of viability we have been overcounting the options. However bear in mind that our task with this work is to count the {\it number of states} in the configuration space of living systems. As per its physics analogue, the configuration space counts not only the states that the system might currently be in; it counts the states that the system could possibly be in, that is the space of {\it all} possibilities. Similarly, we include in our biology count all the ways a biosphere might have evolved or otherwise come to exist, given macroscopically-similar initial conditions.

In summary:
\begin{itemize}
\item The Hilbert space of a quantum many-body theory or quantum field theory, that has a sufficiently-complex spectrum, may have a band of bound states nearly degenerate with the ground state. Such states are every bit and in every way as much a part of the spectrum as the elementary excitations, and hence must be counted when the ground-state degeneracy is assessed.  Such states may very rarely be actualized, and the Universe may be highly non-ergodic with respect to them, but we still count them as possible states in the Hilbert space.
\item When such sectors of states are formed by a self-referential and repeatable process, the number of new states (and hence `degrees of freedom') that may be formed can grow like $2^M$ where $M$ is the number of hierarchical levels of the system, each one contained and concealed within the previous.
\end{itemize}

We choose this quantity, the total number of (micro)states, because it is the quantity that allows us to establish a parallel with Boltzmann's interpretation of entropy in physics, and a parallel with the current estimate for the Universe's entropy. Boltzmann's formula counts all  possible microstates that a system could otherwise have been in, but is not at the present moment. Therefore, in our biological study we count both actual {\it and} potential, that is possible, states.

Additionally, for this space of possibilities we will concern ourselves with the configuration space spanned by the degrees of freedom in the system. In the first examples above these correspond to each base pair, their structured organization in amino acids, genes, unicellular genomes, and finally the resulting multicellular phenotype. If a particular biosphere is predicted to be a possibility by a given configuration of the `degrees of freedom' in the living system, then it must be counted as a possibility spawned by those `degrees of freedom' in our estimation. This is true whether that biosphere is or is not possible to realize by whatever constraints may be imposed on that living system, be it the number of stars, the fitness level, or prospects of extinction.  

Our next task now is to review and develop a key theoretical tool for modeling such systems, which is the Theory of the Adjacent Possible, to which we now turn.

\section{The Theory of the Adjacent Possible}
\label{tapequation}

In summary, what the previous sections indicate is that the configuration space we use in physics, and which is part of the Newtonian paradigm, does not account for the formation of complex structures and the subsequent emergence of new bound states and their dynamical regimes associated with the new emergent structures of biological systems. The conclusion is that in biology we cannot write underlying governing laws, in deterministic form, that can be turned into equations of motion for predicting and explaining the development of biological complexity. At the very least this is a pragmatic impossibility.  Some go further and argue that the biosphere is not subject to any entailing law, even in principle \cite{eros}. For our purposes, we don't have to decide between the pragmatic and in principle objection to wonder if another approach might be helpful.

\subsection{Formulating combinatorial innovation: the TAP equation}

As an alternative way to describe biological systems, one of us (S.A.K.) proposed the Theory of the Adjacent Possible (TAP) to substitute the configuration space \cite{TAP}. The argument states that at any given moment in the evolution of states in the biosphere, it is not possible to prestate what the next step in evolution will be because the underlying laws that determine it (these might be single-site mutations, speciations, extinctions, etc.)\ cannot be predicted in advance. Nevertheless, what happens next is limited by what we already have; evolution expands unpredictably through the space of neighbouring possibilities enabled by combinations of existing items.

The purpose of this section is to compute a finite number for the number of states in the biosphere, $N_{\rm Bio} (t)$. We give one example of a number we found so far. This is obtained via the TAP equation \cite{TAPeqn,TAPeqn2},\footnote{Not to be confused with the Thouless--Anderson--Palmer equation of spin-glass theory.} the behaviour of which we investigate in Section~\ref{solvetap} and more thoroughly in a companion paper \cite{TAPmath}. This  equation (technically a class of closely-related equations) was developed based on combinatorial innovation as a particular version of the Theory of the Adjacent Possible. It can be applied to ideas in society, the invention of patents, human creativity, and also to living cells and organisms \cite{TAPeqn}.\footnote{Inventions, like all human endeavour, are a product of biology. Steel et al.'s results in Ref.~\cite{TAPeqn2} were intended to describe human technology, but each registered patent corresponds in the utmost reductive case to one brain state in a one-to-one mapping. This holds even though in reality an invention is the culmination of a succession of brain states ranging from weeks, months to decades and sometimes a lifetime. Therefore an identification between registered patents to biological states is readily available.} 

It goes like this. We start with the number of states at a given finite time interval. Then we want to know what is possible after this. We consider all possible permutations of subsets and sum them giving each a weight which reduces as the number of included elements becomes larger, as we presume it becomes harder to find and test new combinations. The resulting equation for the number of states $M_t$ at time $t$ looks like this \cite{TAPeqn,TAPeqn2}:\footnote{We have chosen to start the summation from 2, so that separate objects have to be combined to make new ones. Refs.~\cite{TAPeqn,TAPeqn2} instead start the sum at 1, arguing that new objects can also be developed from evolution of a single starting object. Either option is valid depending on context and the differences are marginal.}
\begin{equation}
M_{t+1} = M_t (1-\mu)+ \sum_{i=2}^{M_t} \alpha_i \left( \!\!\! \begin{array}{c} M_t\\ i \end{array}\!\!\! \right) \,.
\label{tap_eq}
\end{equation}
where the $\alpha_i$ are a set of decreasing constants accounting for the increasing difficulty in linking up elements and the final term is the combinatorial combinations of existing elements, while $0\leq\mu<1$ is the extinction/obsolecense rate of previously-created but unsuccessful element combinations. As it stands Eq.~(\ref{tap_eq}) is not quite well-defined due to $M_t$ not being constrained to be an integer, but this can readily be fixed for instance by analytic continuation of factorials using the Gamma function. Alternatively we can make stochastic versions of Eq.~(\ref{tap_eq}) \cite{TAPeqn2}. Once we have chosen these parameters and the elements that are going to be combined (e.g.\ amino acids, or cells, or organisms, etc.) we will have $N_{\rm Bio} (t) \equiv M_t$. An illustrative implementation is carried out in Section~\ref{roadto}. For now let's keep examining the behaviour of combinatorial innovation.

The interesting thing about this equation (even though simplistic and deterministic) is that it has a rate of explosive growth, a rise in the number of states, as time evolves. There is a `time to infinity' timescale at which growth suddenly explodes. In reality it does not diverge at a finite point in time as every term in the summation is finite, but the number of combinations rapidly become unimaginably large as we soon see.

The evolution of human technological development has been observed to exhibit the kind of behaviour described by the TAP equation \cite{TAPeqn}. C.f.\ human science: from 15,000 years ago to 5,000 years ago the diversity of tools developed by {\it Homo sapiens} hardly changed. The number of tools available in that $10,000$-year timespan increased only from about a dozen to a few hundred. In contrast there were only 66 years between the first flight across the Atlantic and the first landing of a spacecraft on the Moon. So human creativity and endeavour has come from a long slow build-up in pre-historic times to a phase of explosive growth in recent decades. 

\subsection{Solving the TAP equation}\label{solvetap}

The TAP equation can be solved analytically under various circumstances, which we fully explore in a different article \cite{TAPmath}. The analysis complements existing investigations by Steel et al.~\cite{TAPeqn2}, who solve variants on the TAP equation by taking a continuum limit, rather than taking the discrete TAP equation itself. As their models do not exactly correspond to Eq.~(\ref{tap_eq}) our results show some differences in detail.

The simplest situation is the case of constant $\alpha$ and $\mu=0$. The behaviour is exactly as expected. Moreover, retaining $M_t$ as the upper limit of the combinatorial sum is shown to be crucial. The key observation is that the sum of combinatorials for constant $\alpha$ is just the sum of the corresponding row of Pascal's triangle minus its first two entries, which is $2^{M_t}-M_t-1$ . Then the TAP equation becomes, 
\begin{equation}
M_{t+1} = M_t (1-\mu) + \alpha \left( 2^{M_t} - M_t -1 \right) \,.
\label{tapeq}
\end{equation} 
Ultimately the second term will always dominate, but not necessarily initially if $\alpha$ is very small .

To see the rapid growth this solution implies, we first set $\alpha$ equal to 1; starting with $M_t = 2$, the sequence goes like this:
\begin{itemize}
\item $t=0$,  $M_t=2$
\item $t=1$, $M_t = 3$
\item $t=2$, $M_t = 7$
\item $t=3$, $M_t = 2^{7} -1 =127$
\item $t=4$, $M_t = 2^{127} -1 \simeq 10^{38}$
\item $t=5$, $M_t \simeq 2^{10^{38}} \simeq e^{10^{38}}$
\end{itemize}
This is remarkable; only one more step is needed to vastly exceed the number representing $\Lambda$'s entropy within our observable Universe. This is consistent with what Steel et al.\ \cite{TAPeqn2} observe. Indeed, in their continuum approach they find the number diverges to infinity in finite time. However solutions to the discrete TAP equation do not do this, because the term being added to the sum is always finite, whatever $t$ is, though they become unimaginably large extremely quickly. Clearly real-world realizations for $\alpha$ cannot be this large or $M_t$ would have escalated in the far-distant past, and there are no such near-infinities observed in Nature.

What is a reasonable value for $\alpha$? It measures a timescale. There are $10^{17}$ seconds since the onset of life on our planet. It appears {\it from real-life applications} that $\alpha$ must be tiny at least in some applications. It took thousands of years for humans to go from having tens of tools to having hundreds of tools, despite there being numerous ways to combine just 10 tools. To make the acquisition that slow, $\alpha$ must be very small even in units of years.

One may wonder why such large numbers occur when using the combinatorial formula, when such numbers are not achieved by standard configuration space permutations of the number of particles. The explanation lies in the concept of `opacity' that arises at each step in the elements at play that are available to be combined into novel elements. Suppose object 3 comes into being from a merger of objects 1 and 2. Object 3 then immediately forgets its origin, which means that at the next time step it is free to combine once again with object 1 or 2 to create yet another novel item.\footnote{This is well illustrated by the Jurassic World films. In the 2015 film the T-Rex and velociraptor genomes are combined to create the fearsome Indominus Rex. By the 2018 film this has been forgotten and the Indominus Rex genome is combined (again) with the velociraptor genome to create the malevolent Indoraptor.} It is this ability to repeatedly recombine that leads to the super-combinatorial group of outcomes; had we excluded this possibility we would not get any innovationally-interesting outcome as the problem would return to being permutational.

We should anticipate slow growth while $t<\alpha^{-1}$, as each step will then be adding (deterministically or probabilistically) a number of order $\alpha$ to $M_t$, which by assumption is much less than $M_t$ itself. Eventually once $\alpha 2^{M_t} \sim M_t$ the second term dominates and the explosion takes place, at a time which is of order $\alpha^{-1}$. Indeed we see this in numerical implementations as well as in our analytic work \cite{TAPmath}.

The extinction parameter $\mu$ is the fractional rate per timestep at which current elements disappear. By comparing the growth and extinction terms, there is always a critical extinction rate which divides the late-time evolutions between explosive growth of $M$ and $M$ dying out to zero. Unless the parameter is tuned to be extremely close to this critical value, it does not have a very significant impact on the evolution and hence its precise value is not of great importance.

The illustrative results that we show later use a version of TAP where $\alpha$ varies with $i$ to suppress multi-element combinations. We adopt the power-law form
\begin{equation}
\label{e:TAPpl}
\alpha_i =\frac{ \alpha}{a^{i-1}} \,,
\end{equation}
where $a=1$ recovers the previous situation. This too is amenable to analytic summation of the combinatorial term, yielding
\begin{equation}
\label{e:planal}
M_{t+1} = M_t (1-\mu) + \alpha a \left[\left( 1+\frac{1}{a} \right)^{M_t}  -\frac{M_t}{a}-1\right] \,.
\end{equation}
In Ref.~\cite{TAPmath} we derive and explore Eq.~(\ref{e:planal}) in detail. Here we simply note that the effect of introducing $a$ is to lengthen the timescale of the blow-up, proportionally to $a$ in the limit $a \gg M_0$, without a qualitative change to the outcome. The normalization $\alpha$ is essentially degenerate with the scale of the timestep (provided the extinction term is suitably rescaled); if for example we choose $\alpha = 0.1$ the shape of the curve is unchanged but it is sampled at ten times the frequency.

\section{The road to organic complexity}
\label{roadto}

We are now finally in a position to develop the main argument of this paper. The sections up to now were necessary stages of reasoning to develop a step-by-step logical line of argument that the TAP equation is a suitable tool to count the statistical ensemble of (living) microstates on the planet, taking into account physics' point of view and methodologies. 

We now want to apply the TAP equation to estimate the size of the configuration space of biological systems, $N_{\rm Bio} (t)$. We will throughout refer to $N_{\rm Bio} (t)$ as a {\bf lower limit} on the real number of distinct microstates existing in the biosphere, which, for reasons explained in the text, we expect to be considerably larger. Our target remains the comparison with the number of cosmological microstates that correspond to the entropy of today's Universe $S_{\Lambda}$. We will want to proceed conservatively to avoid the chance of over-counting of states. 

We remind the reader of the note at the end of the introduction, that when we refer to `entropy' and `states' of biological and other non-equilibrium systems, we are aware that these are not as yet unambiguously defined. The terms should be viewed as analogies, or even as metaphors. 

\subsection{The choice of (out-of-equilibrium) thermodynamical variables}

In statistical mechanics, a microstate is the arrangement of each molecule in the system at a single instant, describing the specific microscopic configurations that it may occupy with a certain probability in the course of its thermal fluctuations, consistent with producing an externally-observable macrostate.  

Then, a macrostate is defined by the macroscopic properties of the system, such as temperature, pressure, volume, etc. The macrostate variables are the things which are observable by whatever apparatus is being deployed.  In this description, microstates appear as different possible ways that the system can achieve a particular macrostate.

We will give an example of how TAP in Eq.~(\ref{e:planal}) can be used to model biological diversity. For the definition of the thermodynamical properties we will consider only one system ${\cal B}$, our biosphere. We will estimate the number of microstates consistent with a macrostate and macroscopic variables, counting the set of microscopic configurations corresponding to all possible alternative lifeforms that could have arisen from the same set of initial conditions.

Defining macroscopic variables is choosing the coarse-graining level at which we wish to examine the system, which is normally governed by the type of experimental apparatus in play. The more deeply we wish to peer into a system, the more complex the macroscopic description. We are going to keep things as simple as possible by staying blind to all `information' except whether the biosphere is living or dead. These are our two macrostates. The `Dead' macrostate is the one of conventional cosmology, as inventoried in Chapter~\ref{inventory}. The `Living' macrostate is the arena where the TAP equation operates.

A microstate $j$ is in the macrostate $\alpha$ if the value of the relevant macroscopic observable for the microstate  is equal to the value of the macroscopic property that defines the macrostate, ${B}_{j}={B}_{\alpha}$. Although the variables describing the microstate do not include the variables describing the macrostate, we assume that as usual the latter may be derived from them, i.e.\ given the complete microscopic description we would be able to say whether the state was Living or Dead.

For our purposes the macroscopic variable labeling the macrostate $\alpha$ is the biological property $B_{\alpha}$, which takes two values: $B_{\alpha}={\cal A}$  for `Alive', i.e.\ a living biosphere, and $B_{\alpha}={\cal D} $ for `Dead', i.e.\ a biosphere with no life.  This requires the following definitions:
\begin{definition}
\label{alive}
We define the macrostate of the biosphere to be `Alive', $B_{\alpha}={\cal A} $, if it contains at least {\it one} living organism.
\end{definition}
\begin{definition}
\label{livingdef}
We define a living organism, following Ref.~\cite{paper2}, as a Kantian Whole within a Type III system that is a non-equilibrium self-reproducing system with a metabolism,  an identity, and a boundary, and which is capable of open-ended evolution by heritable variation and selection or drift.\footnote{The definition of a Kantian Whole can be found in Refs.~\cite{Kbeyond,paper2}.}
\end{definition}
\begin{definition}
The macrostate of the biosphere is `Dead',  $B_{\alpha}={\cal D}$, if it doesn't satisfy the condition to be `Alive'.
\end{definition}

Our argument is unaffected if a different reasonable definition of `alive' were substituted for this choice give by Definition~\ref{livingdef}. Presently there is no consensus on how best to define life, nor even on whether it is in principle possible to do so \cite{cleland}. Those who argue that consciousness lies on a continuum are led to the consequence that life does too. 

We choose to count all microstates corresponding to the set of biospheres ${\cal B}_i$ that could have evolved from the same initial conditions on our planet, under the condition that the biospheres have the macroscopic property $B_{\alpha}={\cal A}$. The `Dead' macrostate corresponds to the conventional calculation of the Universe's entropy, where the state space is fixed and life is ignored. Our existence instead conditions us to be in the `Alive' macrostate. How numerous are its microstates? 

Based on experience with fixed state spaces, one might imagine there have to be many more dead states than alive states, because the atoms that make up a delicate complex living thing can be redistributed in many ways that extinguish its life. Indeed, we argued just that earlier for the putative eukaryote Hamiltonian. But that is not what is happening here, because life has the opportunity to create, and potentially occupy, new states that simply did not exist before. By our assumptions, in a living biosphere the arrangement of the atoms is not a complete specification of the microstate of the living system, hence it is quite possible for living states to dominate.

The numbers of microstates we will derive here are purely indicative and would alter with other functional forms for the TAP equation along with chosen parameter values. Further, a more sophisticated example of TAP portraying the biosphere could require additional parameters and their time evolution. Because the case we chose to treat in the next section is highly simplified, our numbers for $N_{\rm Bio}$ form only lower bounds on the real numbers we believe are produced by complexity in the biosphere. 

Lastly, note that the `degrees of freedom' we choose for $M_0$  need not necessarily form a complete set, nor might they be the most optimal basis on which biological evolution hinges. We have for the current purposes encapsulated the effects of additional parameters and behaviours that could slow down evolution through the choice of a large, and constant, extinction rate to ensure the numbers we present at the end form lower bounds. In all of the functional forms of the TAP equation in the literature \cite{TAPeqn,TAPeqn2}, and the variations we studied in \cite{TAPmath}, the explosive super-growth rate is achieved sooner or later.

\subsection{Our complexifying biosphere}

Life in our biosphere is assembled from a `building block' set of six atoms known as CHNOPS (carbon--hydrogen--nitrogen--oxygen--phosphorus--sulpher). CHNOPS represents the set of six chemical elements from which most of the biological molecules on Earth are constituted. Throughout the evolution of life on the planet covalent combinations of atoms in these molecules evolved to form ever more complex compounds of CHNOPS elements. 

Due to the infall of meteorites, and local biosynthesis, the Hadean earth had a considerable diversity of small molecules, including amino acids which are the building blocks of proteins, lipids, and sugars. The formation of nucleotides was rare though present \cite{deamer}. Two sites of further synthesis are often considered: hot thermal undersea vents \cite{NickLane}, and terrestrial tidal or fresh water pools subject to wet/dry evaporation cycles. Such cycles can drive polymer formation of larger peptides and polynucleotides \cite{damerdeamer}. From such venues, early molecular reproduction presumably arose.\footnote{There are two competing views on how this might have taken place which are of some relevance to us. In one, molecular replication arose by a `nude' RNA molecule learning to template replicate itself, though experimental work to achieve such a molecule has yet to succeed. In this view life did not arise until the emergence of long single- or double-stranded RNA sequences, though ribonucleotides are rare on the early Earth \cite{deamer}. The second view is that molecular reproduction first arose by a transition to collective autocatalysis in sufficiently-rich chemical soups \cite{KauffFarmer}. Collectively-autocatalytic DNA sets, RNA sets, and peptide sets have been constructed experimentally, the last showing that polynucleotides are not essential for molecular reproduction. Such sets have been found in primitive bacteria and archaea, and overlap in a way suggesting a smaller, yet still autocatalytic, set was present before archaea and bacteria diverged \cite{Xavier}.}

Evolution of amino acid molecules led eventually to nucleic acids (which hold together RNA and DNA on a backbone of sugar and phosphates) and to the ATP molecule which powers the cell, as well as any organic molecules needed for cell function. For these reasons we will consider the evolution of organic compounds of CHNOPS, and count the complexity that could have instead arisen from the same initial chemical elements in place of our own. This means counting the number of all alternative sets of possible molecular compounds which could have formed from those same 6 initial elements. 
 
This includes all alternative composites of organic molecules, from alternatives to the simplest glucose molecule, all the way to alternative versions to the dominant species on the planet, ourselves, while encompassing the vastness of possible biological diversity that would have led to the rise of alternative prevalent species. 
 
This is not a simple task as we obviously don't have an algorithm to predict what possible complex life forms could have formed instead of our own. To that end, the TAP equation, with its accounting of combinatorial innovation (counting possible combinations, and combinations made of combinations), proves to be the ideal phenomenological tool. We will illustrate using a particular version of the TAP equation from the previous section with a choice of parameters that strongly suppresses innovations from more than two originators, to account for the difficulty in successfully interacting more than two items simultaneously; other versions of TAP are being developed and will also merit exploration, initial steps of which we carry out in Ref.~\cite{TAPmath}.

The full set of chemical elements in CHNOPS were first made available in the cosmos by supernovae ejection from Population-III stars very early on, possibly at redshifts as high as $z\sim8-10$ \cite{TypeIII-1} which using the {\it Planck} satellite collaboration's best-fit cosmology \cite{Planck} corresponds to a look-back time of 13.15 to 13.31 billion years (the age of the Universe for the best-fit parameters is 13.78 billion years). It took the Universe only about 50 million years from the Big Bang to produce the building blocks of life. These findings are supported by the observation of heavy-element emission lines --- elements which could not have been created in the Big Bang --- in quasar emission spectra~\cite{quasaremission}.

We know that well before the formation of our Solar System around 5 billion years ago, amino acids were abundant in meteorites. In fact the most abundant type of asteroids at that time are carbonaceous asteroids rich in CHON elements. Further, the 7-billion-year-old Murchison meteorite was found to have of order $10^5$ different organic molecules \cite{StuOrgMol} and around 96 different types of amino acid \cite{96aminoacids}.

Now the road of growing complexity of organic molecules continues, and next is the formation of the first proteins on Earth around 3.9 to 3.5 billion years ago (cosmic redshifts $z\simeq 0.34$ to $0.30$). These were simple peptides, forming out of a set of only 10 prebiotic amino acids \cite{prebioticaminoacid}. These pre-proteins were produced by simple chemical processes, but contained the requisite information to produce complex folded proteins. Folded proteins are required to produce the precise structures which are essential for the functions that support life. These early proteins were made in auto-catalytic processes and were not yet encoded in the long chains of nucleic acids in RNA, as will later be the case. 

The next milestone in organic complexity was the completion of the map of amino acids which make up the building blocks for life on Earth. This comprises 22 proteinogenic amino acids,\footnote{The 20 standard amino acids of the genetic code plus 2 extra ones for which there are special coding sequences.} selected by Nature through evolutionary pressure, to add on to the initial prebiotic 10. 

From here on the production of ever more complex macroproteins required the encoding of their long strands of polymers in nucleic acids to allow for the accurate replication and precise folding of its complex polypeptide chains.  This marked the appearance of the first proteinogenic proteins and this became possible due to the arrival of a macromolecule storing genetic information --- the self-replicating RNA molecule. From here onwards ever more complex proteins were synthesised by (proto-)ribosomes which read out their code from a strand template in RNA nucleotide chains \cite{rnaworld}.

We have landed in the RNA world and the history of life will never be the same. The biosphere has learnt to store information and how to remake its creations from it.

We have also reached that which may be the common ancestor to all known living things. Every organism on the planet today can trace their lineage back to a global ancestor. The comparison of genomes indicate that the last universal ancestor common to all living beings is the RNA polymerase, coding up proteins in its base pairs made up of the same 20 amino acids.

We pause here to take stock of the complexity of life on the planet at this time, which we will call the time of {\it Template Synthesis}, $t_{\rm TS}$, and corresponds to roughly 3.5 billion years ago. We will try to estimate how the configuration space of all possible organic molecules looks around $t_{\rm TS}$, the time that RNA appeared on the planet, and assess if the amount of {\it `information'} on the planet at this point already exceeded that contained in $\Lambda$.

We also pause here to cross over back to cosmology and take a look at what was happening there.  For, at the very same time that RNA our common ancestor first made its appearance around 3.5 billion years ago, something quite unique was happening everywhere in the cosmos. 3.5 billion years ago corresponds, if we take {\it Planck} satellite's best-fit parameters, to redshift of $z\sim0.3$. This is the time of matter--dark energy equality, and marks the epoch in which dark energy came to lead the energy budget of the cosmos, on its way to its clear domination of the total density today.

The mystery of the appearance of the dark energy material and the study of this intriguing type of energy dominates modern cosmology. In particular we wonder why it came to dominate only recently in the history of the cosmos and not before. This is called the problem of coincidence. We are here identifying a second coincidence: that something equally relevant in the bio-history of the planet is taking place at the same time that dark energy comes to dominate. We speculate on a possible meaning to this coincidence in Section~\ref{darkenergy}.

\subsection{TAP model biospheres}

Having arrived at analytical expressions which represent the behaviour of the TAP equation in the previous section, we must choose parameter values which will best mimic the evolution of the system in consideration. As the behaviour of the TAP equation is fairly generic, the key issue is to have well-motivated choices which can later be refined in more detailed explorations of parameter space.

We return to the estimation of a lower limit on $N_{\rm Bio}(t_{\rm TS})$ which we choose to correspond to organic compounds made out of the initial six elements present in CHNOPS at the time of the first RNA molecule. That is the first parameter in the TAP equation~(\ref{tap_eq}), $M_0=6$. We bear in mind that our goal is the comparison of $N_{\rm Bio}(t_{\rm TS})$  with the number of microstates that correspond to $S_{\Lambda}$, which is $N_{\Lambda}=\exp(10^{124})$. Once again we note that the degree of diversity in the biosphere is ever increasing, so if we find $N_{\Lambda} < N_{\rm Bio}(t_{\rm TS})$  at $t=t_{\rm TS}$ then we will necessarily have $N_{\Lambda}\ll  N_{\rm Bio}(t_{\rm today})$.

We use the TAP equation with a power-law fall-off of $\alpha_i$, as in Eq.~(\ref{e:TAPpl}). This is necessary to suppress overly complex multi-element combinations at least initially, as observed in primitive simple-to-complex organic molecules evolution \cite{Oparin-Haldane-Miller-Urey}. 

The normalisation parameter $\alpha$ is degenerate with the timestepping (provided the extinction is suitably rescaled) and simply decides how many outputs TAP gives between each increment of time; we choose $\alpha = 0.1$ for illustrative purposes.

There is no reason to presume that the length of the timesteps, expressed in say years, remains the same throughout the evolution. Indeed one should expect all parameters of TAP to be subject to revision as the system ascends the hierarchy of complexity. Plausibly the later steps are longer as it becomes harder to successfully merge two or more highly-complex systems to create something novel. For example initially the rate of formation of new organic molecules is high, since there are many simple organic molecules in the environment and it is easy for them to find each other and create new compounds. Towards the end as more and more complex molecules have now formed; these will be less abundant in numbers compared to the more simple molecules and their structure will need to be encoded in RNA for reliable reproduction.

For guidance we can look to evolutionary timescales for RNA polymerases during the present epoch. Their development has become synonymous with the success of the species that they code for; to sustain and multiply themselves the molecules must code for species which have selective evolutionary advantage. The timescales for species evolution remain controversial, but it has been suggested that it requires about a million years for major evolutionary changes to become consolidated \cite{Uyeda,Gingerich}, even though short-term evolutionary processes may take place much more quickly. With $\alpha = 0.1$, we can thus consider each step to be of order $10^5$ years.

The background extinction rate (i.e.\ the rate that pertained before humanity's drastic effect on the world's habitat systems, the so-called Anthropocene) is estimated at roughly one species per million species per year \cite{May}. Consistency with a complexifying biosphere of course implies TAP's extinction must be at a sub-critical rate; after guaranteeing that, its precise value has little impact \cite{TAPmath}. We find of order $10^{-5}$ to $10^{-4}$ to be suitable values for $\mu$ (which measures the fractional extinction per timestep).

The longest proteins occurring in Nature today have order 27,000 to 35,000 amino acids,\footnote{Titin is amongst the most frequently-occurring proteins in muscle tissue, and with this length is the longest known protein.} but those present 3.5 billion years ago would have been much shorter. For this reason we have included the power-law parameter $a$ which suppresses the formation of proteins of more than a given length. According to Ref.~\cite{Tiessen}, over a wide range of species the most frequently-occurring proteins have a length of a few hundred amino acids. So in Eq.~(\ref{e:TAPpl}) we will choose $a$ to favor organic molecules which have formed proteins up to a length of 250 amino acids, and act as a slow suppressant of proteins longer than that. A lognormal distribution for protein size gives a fit suppression value for $a$ of two to three hundred.

For our example here we choose a simulation based on the basic constituents of CHNOPS, $M_0=6$. We evolve TAP in Eq.~(\ref{tapeq}) since the onset of life on the planet, $t_{\rm OL}=$ 3.7 billion years ago, up until the time of template synthesis, $t_{\rm TS}=$ 3.5 billion years ago. We obtain therefore the number of different molecule composites that could have been created by biology at that time, given by $N_{\rm Bio}(t_{\rm TS})$. Again we emphasise here that given the simplifications of the model we are treating here, the value of $N_{\rm Bio}(t_{\rm TS})$ estimated is a lower limit on the parameter's real value in the biosphere.

\begin{figure}[t]
\centering
\includegraphics[scale=0.5]{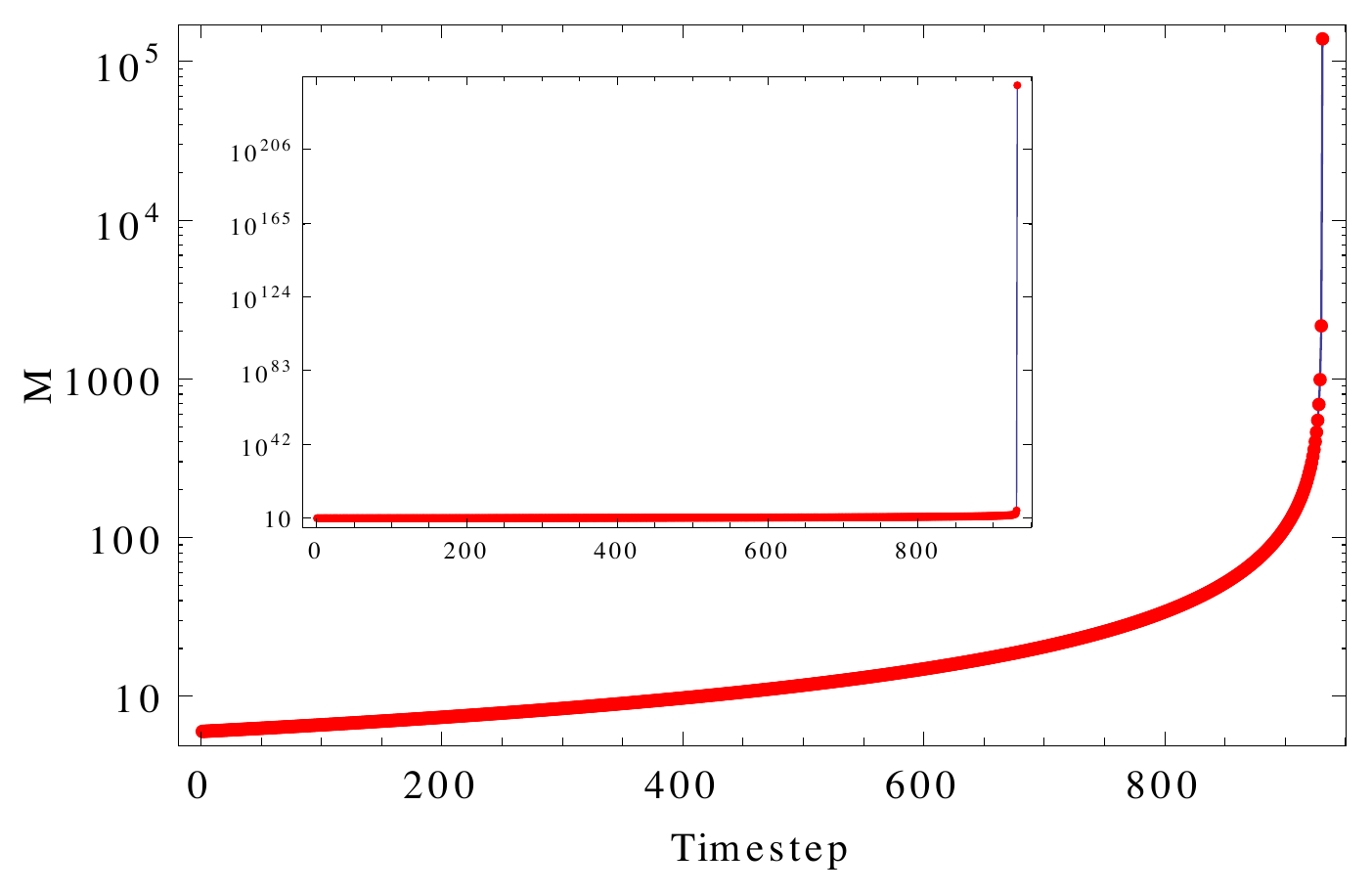} 
\caption{We obtain the characteristic hockey-stick behaviour with example evolution $M_0 = 6$, $a=250$, $\mu = 10^{-4}$, $\alpha = 0.1$. The insert simply shows one further point, $10^{240}$ (the last before machine overflow) whose inclusion drastically rescales the $y$-axis. One further point given by applying Eq.(~\ref{e:planal}) takes us to $10^{10^{237}}$, hence beyond $N_{\Lambda}$ and achieving the goal of this work: to demonstrate that there are more microstates contained in the phase space of the biosphere than there are in that of the entire remaining Universe. }
\label{M06}
\end{figure}

We obtain the characteristic hockey-stick behaviour of Steel et al.~\cite{TAPeqn2}, shown in Fig.~\ref{M06}. Growth starts slowly due to extinction and multi-element suppression rates. The last point we obtain in the plot before machine over-flow is $10^{240}$. Equation~(\ref{e:planal}) allows us to compute the point after this one which already supplants $N_{\Lambda}$, $N_{\rm Bio}(t_{\rm TS})\sim10^{10^{237}}$.  

We cannot stress enough that a different choice of parameter values in Eq.~(\ref{e:planal}), or possibly time-dependent and higher-dimensional parametrizations of the TAP equation, would change the microstates estimate for the biosphere we found. Different runs we have carried out, choosing different parameter values, yield both much  larger and much smaller final values for $N_{\rm Bio}(t_{\rm TS})$ though the hockey-stick divergence is generic. $M_0=6$ cannot be varied easily, since the ranges we found for the remainder of parameters $\mu$, $a$, and $\alpha$ are anchored directly on the choice of $M_0$ as the 6 CHNOPS atoms, which represent the  `elementary particles' for the simulation.  The values of $\mu$, $a$, and $\alpha$  were subsequently extracted, under the CHNOPS assumption for $M_0$, from a body of biological data on the evolution of complex organic molecules. So to vary $M_0$ would require specifying a different fundamental unit for TAP in Eq.~(\ref{e:planal}) and a complete reevaluation of the biological phenomena that each variable parametrizes, as well as the the evolution of the organic compounds TAP is describing.

However even with the high extinction rates we used, together with a relatively strict suppression of the likelihood for the formation of large, multi-element compounds, the power of combinatorial innovation of the TAP mechanism in counting biological complexity from first principles usually wins and manifests itself in the observed late-time blow-up. This completes our goal with this work. We conclude that by the time of template synthesis, $t_{\rm TS}$ the size of configuration space in our biosphere, vastly exceeded the size of the configuration space of the Universe stored in $\Lambda$,
\begin{equation}
N_{\Lambda}(t_{\rm TS})\sim 10^{10^{120}} \ll N_{\rm Bio}(t_{\rm TS})\sim 10^{10^{237}} \,.
\end{equation}
Note that at that time of {\it template synthesis}, $t_{\rm TS}\sim 3.5$ billion years look-back time, which corresponds to $z\sim 0.3$ (according to Planck's best-fit parameter values) the universe has already begun accelerating. Consequently, the vacuum contribution to the cosmic energy budget at that time is already established at something like its current value. 

From then on until today it has only continued to inflate unimaginably, even if punctuated at times by mass extinction events, which potentially regulate TAP's large numbers. Apart from those, that is precisely the action of the TAP equation at each timestep in its late-time evolution.

\subsection{Summary: A status checkpoint so far}

We have performed a first attempt to account numerically for the growth of complexity in life on the planet, starting from the onset of life up till the time of template synthesis --- the appearance of the first RNA molecule.

As physicists we naturally want to undertake such a counting referring to well-defined physical quantities in a familiar language, as is the case of statistical mechanics. To that end we expressed complexity in terms of the number of microphysical states that assemble into living systems. To encompass living microstates we only needed to extend the traditional concept of microstate as a specific microscopic configuration of molecules to encompass also states formed by the emergence of novel, underivable, bound states. This underivable emergence is, as we saw in this section, a defining characteristic of organic evolution as it flourishes. We maintain the statistical-mechanics definition that the configuration space is formed by the ensemble of all possible microstates contained in living systems, and bestows on the biosphere its global variables and macrophysical properties as a whole.

It is no coincidence that we have chosen the time of the first template synthesis in the evolution of life on Earth to portray an example of the theory of the adjacent possible, and the TAP equation, at work.  The appearance of the first RNA polymerase is the moment that life first began to replicate itself through an ingenious and sophisticated new process. Life learnt that if it stored the information to replicate itself in long chains of nucleic acids, it could then synthesize copies of itself using ribosomes which read out this code. The replicas produced this way were much more accurate and exact, and they allowed in turn that the precious adaptive mutations that one generation had painfully learnt in the survival game of living and dying would not to be lost to the next generation. Subsequently, the accurately-replicated new offspring arose stronger and better adapted to the ever-evolving environment than previous generations. This allowed an unprecedented explosion in the adaptive power of rapidly-evolving organisms. 
 
This defining event in the history of life altered, forever and unrecognisably, the face of the living planet with regards to the luxuriant array of diversity and complexity of organic structures that developed here onwards. Such a massive flourishing of possibilities can only be encapsulated in the super-exponential, hockey-stick growth which is distinctive of TAP. 

In the remainder of life's history on Earth there were five near-global mass extinction events and threats to the survival of life on the planet, perhaps six if we include the current Anthropocene \cite{anthropocene}. However even if the current extinction epoch includes that of humans ourselves, Nature will surely reclaim our abandoned planet, perhaps this time starting with insects, given their fast reproductive and adaptation rates \cite{afterhumans}. As the saying goes, {\it life finds a way}. To this we are adding that it finds, sooner or later, a way to TAP.

\section{Biological dark energy}
\label{darkenergy}

We end with a certainly outrageous, yet in some ways compelling, speculation. In this article our main argument has been that the biosphere cannot necessarily be neglected in an accounting of the number of states in the Universe, due to emergent complexity. If indeed the Universe's `entropy' is much larger than previously thought for this reason, it does not immediately lead to any change in our understanding of how the Universe evolves; recall that the discovery that black holes or the cosmic horizon dominate the entropy from physical sources did not prompt any change to the standard cosmology. Rather, our results are a tool to assess the specialness of initial conditions. Nevertheless, it is tempting to speculate whether there can be a dynamical effect, prompted by the coincidence of dark energy coming to dominate the Universe at almost the same time that life first emerged on Earth. Could the emergence of life have inadvertently caused the emergence of dark energy, in what one might call the ultimate environmental catastrophe?

With standard cosmological parameter values, life's emergence on Earth at around 3.7 billion years ago was at redshift 0.33, while the cosmological constant matched the matter density at redshift 0.30. This is certainly a tighter confidence than the usual `why is Lambda dominating now', though still very plausibly just a coincidence. So could the `information' associated with life's emergent complexity modify the cosmic vacuum energy? 

It is easy to find objections. Firstly, observations show the effects of dark energy at higher redshifts while it was still subdominant, at epochs before the Earth's formation, so we will need to invoke earlier emergence of life at other locations. This is however quite plausible given that the Universe's star-formation rate peaked around 10 billion years ago \cite{SFR}. Secondly, since life emerges the density of dark energy would be expected to be growing with time, whereas observations show the dark energy density to be at least nearly constant in recent epochs (though admittedly only fairly weakly constraining growing, i.e.\ phantom, dark energy models). Thirdly, it is unclear how life emerging at one location could spark a dark energy phenomenon throughout the (observable) Universe, so life would need to be widespread. 

Nevertheless, the logic of this article suggests that the possibility merits further exploration, for instance in case the super-explosive growth indicated by the TAP equation can equate to a single `switch-on' of dark energy which would be consistent with current observations if it happened early enough. Perhaps also there is some connection to the intermittent attempts made to connect dark energy to information and its erasure via the Landauer principle \cite{DEland,DEland2} or entanglement entropy \cite{Capent}. For example Ref.~\cite{DEland} ties the emergence of dark energy to information erasure by star formation, which is a step towards a link with biology. These ideas may also ultimately connect to the holographic dark energy picture \cite{HDE}, the causal entropic principle of Bousso et al.~\cite{causalent}, and the general relativistic entropic force model \cite{entforce}.

\section{Overview}

No one knows what is the space of `degrees of freedom' that natural selection began exploring when life set out on our planet 3.7 billion years ago. We are making the first steps in assessing the consequences of this. We have used the TAP equation to model the biosphere's emerging innovations and complexity, estimating the configuration space of living matter on Earth up until the emergence of the genetic code and of the translation apparatus. 

\subsection{The arguments}

We make two key arguments in this paper:
\begin{enumerate}
\item The processes leading to the appearance and evolution of living systems are radically different from the processes ruling physical systems in which life is absent. This difference is closely tied to extreme non-ergodicity, found only in the phase space of these living systems, and absent when life is not present. 
\item The non-ergodicity occurs because the phase space or Hilbert space of such systems is continually expanding while life is present. This expansion takes the form of a continuous emergence of new bound states of living systems. Because of this perpetual creation of new living composites, our generic physical postulates --- like Liouville evolution in $6N$-dimensional phase space, and the Hamiltonian or unitary evolution of quantum states --- cease to apply in biology. We cannot find the biological equivalent for such postulates. 
\end{enumerate}
The simplest way to illustrate what it means for novel bound states to be perpetually emerging in biology is to imagine that new elementary particle species were continuously and unpredictably forming in our Standard Model of particle physics. How could we ever hope to derive its corresponding Lagrangian?

The same holds for standard microphysical laws, which become insufficient to explain the evolution of living systems in the many-body and long-time regimes. If we wish to extend our theoretical physics understanding of Nature to encompass the description of living systems from an elementary and fundamental viewpoint coming from first principles, us theoretical physicists need to extend our toolkit to include phenomena arising through combinatorial innovation. One of these cases is the theory of the adjacent possible, which we described extensively.

We must highlight the difference between the ergodic and non-ergodic regimes as crucial and key to the derivation of these conclusions. While we are in the ergodic regime, objects can be repeatedly made simply because the conditions to make them occur repeatedly. For instance, in the early stages of molecular chemistry the same small molecules are independently created many times through collisions of their constituents --- the Universe has no need to learn and remember how to make those molecules. The present non-ergodic Universe is radically different, because to create biologically (or technologically) complex items repeatedly requires the prescription for their creation to be learned, stored, and then executed. It is this latter regime which lies outside the realm of the conventional physics description, and for which TAP provides a phenomenological model.

\subsection{The conclusions}

Some of the important conclusions drawn from this merger of cosmology and biology can be summarised as follows:
\begin{itemize}
\item In a theory that is well described by an underlying fundamental model, like the Standard Model of particle physics, the global space of states, the Hilbert space, does not expand. The states that the Universe goes through are vector states which were present in the model from the beginning, and do not appear unpredictably as the Universe undergoes different phase transitions. Whether or not those states are available to the Universe at a given time depends on the temperature regime the Universe is in. All states in the evolution of the Universe along with their effective field theory can be derived from the Standard Model (or its appropriate extension). 
\item In a system that has {\it no} standard model, such as biology, the configuration space genuinely and {\it unpredictably} expands in real time. As the system evolves, new microstates are (combinatorially) found and tested by the system's basic constituents. Those states are genuinely novel, in that they could not have been derived {\it a priori} by any underlying theory. Therein lies the crucial distinction between physics and biology.
\end{itemize}

The central question for cosmology that we are addressing here is:  Does the Universe create large amounts of `information' as it evolves from early times to late  times? We can compare the probable answers to this question as given by two paradigms for cosmological theory. These are:
\begin{enumerate}
\item The Newtonian paradigm (defined above), which asserts that there is a fixed space of states on which fixed laws operate.  
 \item An evolving-laws paradigm, which denies that there are either a fixed set of states or fixed laws.
\end{enumerate}
Different authors, including the present ones, formulate evolving-law paradigms differently,  using various frameworks.  But we all agree on the denial of the Newtonian paradigm, which is sufficient to reach a firm conclusion on the forementioned question.  That is, our differences matter much less than what we agree on.   

We can all agree on the following.  Both paradigms have well-defined quantities called $\{ \rho \}_{{\rm initial\;lawful}}$, the set of `lawful' possible initial states of the Universe of which one, $\rho_{\rm initial}$, was the actual initial state of the Universe, and $\{ \rho \}_{\rm late\;lawful}$, a set of `lawful' final or late states of the Universe of which one, $\rho_{\rm final}$, will be the actual final or late state of the Universe.  Different theories may possess either a final state or a late state, the latter being one after everything interesting has happened. 

There are various options for speaking of measures on sets of states. Some, like the Boltzmann or von Neumann entropies, depend on the theory and only make sense in the first of the two paradigms we are comparing.  But Bateson information, defined informally as `the difference that makes a difference' \cite{Bateson}, is universal enough to let us compare the two paradigms.

Let $I_{\rm initial}^{\rm NP}$ be the amount of Bateson information needed to select the actual initial state of the universe from the possible lawful initial states in the Newtonian paradigm. Let $Q^{\rm NP} (\rho_{\rm initial}^{\rm NP} \rightarrow  \rho_{\rm final}^{\rm NP})$ be the information needed to predict the exact final state given the exact initial state in the Newtonian paradigm. The same quantities labeled with `EL' arise in the evolving-laws paradigm. 

Our key question is, does the Universe learn during its evolution?  In other words is $Q$ order unity or huge? Most cosmologists focus on the $I$,  which is called the `entropy' of the initial state, or just the `entropy' of the Universe.  We instead are interested in the $Q$, which tell us how much information the Universe {\it learns} during its evolution.   This is information that is not present in the initial conditions and also is not a consequence of the actual laws. 

If you believe in the Newtonian paradigm, you must believe $Q^{\rm NP} = 1$, as all the information needed to precisely determine the final state is present in the initial state. But an evolving-laws framework makes possible a creative universe in which $Q^{\rm EL} \gg 1$.  In other words, all evolving-law cosmologists agree that the Universe creates information.  We call $Q$ the creative potential of the Universe.  A biological universe can create more `information' than was present at the Big Bang by conventional accounts.

\subsection{The outcome}

To summarise, we have taken the first steps in bringing the biosphere into cosmology by proposing a tool for estimating the number of classical microscopic states contained in the biosphere configuration space. Within the context of an emergent and non-reductionist view of complex systems, we argued that it is not self-evident that the contribution of the biosphere is sub-dominant to the enormous numbers coming from gravitational physics. Emergent complexity may dramatically expand the available volume of configuration space. As a demonstration, we have adapted equations emerging from the Theory of the Adjacent Possible \cite{TAP}, which show an explosive growth that can be super-combinatorial and hence in principle able to overcome  the traditional permutational accounting of particle configuration space. Living biospheres may be the dominant source of `information' in the Universe! How and whether the potential associated divergence of complexity can be tamed (either in equations or in reality itself) remains to be seen. An exciting speculation is that cosmic acceleration may have a role to play in this.

Has our new result worsened the problem of cosmological initial conditions? Surprisingly, probably not. The vast numbers generated via the biospheres are due to the growth in state space enabled by the emergent complexity of life. Those states simply did not exist until life came along to make them. The newborn Universe had no opportunity to occupy them, nor pays any cost in failing to do so.
 
To end we note that one should not regard TAP as describing actual physical collisions of two or more objects to make a single new one; it operates at a much more abstract and elevated level. It quantifies the {\it remembered} ability of a system to make and remake an object.  The word remembered is crucial; it is not enough to make something once, we must be able to do so again and again. Its existence must be encoded in some way; embedded in RNA perhaps, or passed on through folklore, or written down in a stored patent application. This requirement can become incredibly complex. To make one new leopard we need both the actual parent leopards and the information encoded in their genomes (which comes conveniently packaged within the leopards themselves). But to keep making new leopards requires the entire sustainable ecosystem in which they are embedded. Consider John Muir's famous quote \cite{Muir}: ``When we try to pick out anything by itself we find that it is bound fast by a thousand invisible cords that cannot be broken, to everything in the universe.'' 

\section*{Acknowledgments}
We would first and foremost like to acknowledge Barbara Drossel, without whom several of the connections in this work would never have been made. We thank Niayesh Afshordi, Priyal Bordia, Latham Boyle, Hal Haggard, Wim Hordijk, Jaron Lanier, Roberto Mangabeira Unger, Pierre Martin-Dussaud, Mark Neyrinck, Roger Penrose, James Rosindell, and Carlo Rovelli for discussions.

This research was supported in part by Perimeter Institute for Theoretical Physics. Research at Perimeter Institute is supported by the Government of Canada through Industry Canada and by the Province of Ontario through the Ministry of Research and Innovation. This research was also partly supported by grants from NSERC and FQXi. This work was supported by the Funda\c{c}\~{a}o para a Ci\^encia e a Tecnologia (FCT) through the research grants UIDB/04434/2020 and UIDP/04434/2020. M.C.\ acknowledges support from the FCT through grant SFRH/BPD/111010/2015 and the Investigador FCT Contract No.\ CEECIND/02581/2018 and POPH/FSE (EC). A.R.L.\ acknowledges support from the FCT through the Investigador FCT Contract No.\ CEECIND/02854/2017 and POPH/FSE (EC). M.C.\ and A.R.L.\ are supported by the FCT through the research project EXPL/FIS-AST/1418/2021. We are especially thankful to the John Templeton Foundation for their generous support of this project.

\end{document}